# Analytical Queries for Unstructured Data

**Daniel Kang**

University of Illinois Urbana-Champaign

ddkang@illinois.edu

**now**

the essence of knowledge

Boston — Delft

# Contents

# Analytical Queries for Unstructured Data

Daniel Kang

*University of Illinois Urbana-Champaign, USA; ddkang@illinois.edu*

ABSTRACT

Unstructured data, in the form of text, images, video, and audio, is produced at exponentially higher rates. In tandem, machine learning (ML) methods have become increasingly powerful at analyzing unstructured data. Modern ML methods can now detect objects in images, understand actions in videos, and even classify complex legal texts based on legal intent. Combined, these trends make it increasingly feasible for analysts and researchers to automatically understand the "real world." However, there are major challenges in deploying these techniques: 1) executing queries efficiently given the expense of ML methods, 2) expressing queries over bespoke forms of data, and 3) handling errors in ML methods.

In this monograph, we discuss challenges and advances in data management systems for unstructured data using ML, with a particular focus on video analytics. Using ML to answer queries introduces new challenges. First, even turning user intent into queries can be challenging: it is not obvious how to express a query of the form "select instances of cars turning left." Second, ML models can be *orders of magnitude more expensive* compared processing traditional structured data. Third, ML models and the methods to accelerate analytics with ML models can be error-prone.

Recent work in the data management community has aimed to address all of these challenges. Users can now express queries via user-defined functions, opaquely through standard structured schemas, and even by providing examples. Given a query, recent work focuses on optimizing queries by approximating expensive "gold" methods with varying levels of guarantees. Finally, to handle errors in ML models, recent work has focused on applying outlier and drift detection to data analytics with ML.

# 1

## Introduction

Data volumes are exponentially growing. The bulk of this data is *unstructured data*, or data where the information of interest does not natively conform to a schema. Unstructured data is primarily in the from of text, images, video, or audio. As an example of the massive growth of data volumes, the Tesla fleet of vehicles *alone* produces *exabytes* of data per day. Video data further accounts for over 88% of internet traffic. Beyond video, there are enormous volumes of PDFs (with text), audio (such as podcasts), and images (such as on social media).

In tandem with the growth of unstructured data, our ability to extract semantic, structured information via machine learning (ML) from this unstructured data has improved rapidly. Recently developed large language models (LLMs) like GPT-4 (OpenAI, 2023) can automatically perform complex tasks, such as classifying whether or not complex legal texts invoke textualism as the legal reasoning behind an opinion (Choi, 2023). Cloud providers, ranging from Google to Amazon Web Services, now offer powerful APIs to extract object types and positions from visual data (Ren *et al.*, 2015). Automatic audio transcription services are now fast, accurate, and cheap (Radford *et al.*, 2023).

The rise of these automatic ML methods and APIs has enabled non-ML experts to be able to analyze these increasing amounts of data that reflect the "real world." Legal scholars have used ML to assist in understanding how judges apply legal reasoning across large corpora of case opinions (Choi, 2023), urban planners can use cheap cameras to make informed decisions about city planning (Brisotto *et al.*, 2024), and business analysts can understand customer reviews and automatically classify issues with their products.

Unfortunately, deploying these ML methods automatically remains challenging.

First, even specifying what information to extract can be challenging. An urban planner may be interested in counting the number of times that a car suddenly decelerates in an intersection. But what does it mean to "suddenly decelerate?" Or to be "in the intersection?" Allowing non-experts to define precise query semantics from higher-level goals is a major impediment to the wider adoption of these tools.

Second, using ML for queries can be orders of magnitude more expensive compared to standard structured data queries. GPT-3, a last-generation LLM, takes approximately *350 billion* floating-point operations per input token of text (Kaplan *et al.*, 2020). In contrast, a string comparison may take as few as 20 instructions per input token of text, a *ten orders of magnitude difference.* Aside from computational resources, using APIs can also be incredibly costly. Naively using the Google Cloud Vision API to analyze a small town's worth of video (10 cameras for one year) would cost over *$10 million dollars.* These costs are prohibitive for all but the largest and well-funded organizations.

Third, ML models can silently produce errors. In the urban planning example, an object detection model can fail to identify cars. Worse, the failures can be correlated with features of interest, such as the inclement weather patterns (Suprem *et al.*, 2020). In text, a class of these errors are commonly referred to as "hallucinations."

Although these challenges make deploying ML for unstructured data queries difficult, they have also led to a wealth of research in data management over the past several years. This research spans abstractions, systems, and algorithms for specifying unstructured data queries, answering them efficiently, and finding when errors may be occurring.

To specify queries, researchers have designed interfaces and abstractions ranging from using ML as user-defined functions (UDFs), pure

SQL, new domain-specific languages (DSLs), to exemplar-based. For simple ML-based queries, systems that use UDFs and pure SQL have shown remarkable success, collecting thousands of stars on GitHub. These ideas have also been integrated into commercial offerings, such as DataBrick's hosted Spark. Unfortunately, for more complex queries, no single abstraction or system has been widely adapted. Many research prototypes have been built and deployed with specific use cases in mind, but have not been shown to generalize.

To accelerate and reduce the cost of queries, researchers have developed sampling algorithms, designed systems to leverage accelerator hardware effectively, and tune system configurations for high performance. Collectively, these optimizations can reduce the cost of certain queries by orders of magnitude, which has made queries feasible at scale. Despite these advances, ML-based queries remain incredibly expensive, especially when leveraging state-of-the-art models.

To handle errors in ML models, recent work has focused on detecting errors in ML models via signals from domain experts, detecting errors in training data used to train ML models, detecting model drift, and guarantee accuracy against human performance. Nonetheless, many challenges remain. These tools are often standalone and not integrated into larger unstructured data analytics systems. As a result, handling errors remains an open problem, especially in end-to-end systems.

As we have seen, unstructured data queries have great potential to enable domain experts to understand the real world. Existing research has pushed forward the capabilities of data management systems for this unstructured data, but much work remains to be done.

In the remainder of this monograph, we will discuss the aforementioned research for unstructured data queries and discuss open challenges. Sections 2 and 3 provide the background on unstructured data systems and a high-level architecture of an unstructured data system. Sections 4, 5, 6, 7, and 8 discuss components of the architecture: expressing queries, general query optimization, query optimization algorithms for approximate queries, indexing and storage, and query execution respectively. Section 9 discusses queries over video data and Section 11 discusses open challenges.

# 2

# Background

In this section, we provide background on unstructured data, use cases for unstructured data, modern trends in machine learning, and the relevant materials related to "traditional" structured data management systems.

At a high level, many of the challenges that arise in unstructured data systems are also present in structured data systems. How do we store data and build indexes to accelerate downstream queries efficiently? What query optimization methods should we use? How should uses express queries?

However, the extreme expense of ML/AI methods and the ambiguity around many unstructured data queries give rise to unique challenges for unstructured data systems. We now discuss the background relevant necessary to understand these unique challenges.

## 2.1 Unstructured Data

In this monograph, we refer to *unstructured data* as data where the semantic information of interest is not natively present. In nearly all cases, unstructured data is either sensor data (images, video, audio, LIDAR point clouds, etc.) or text. We focus on *semantic* information, such as object types and positions in an image.

Sensor data now consists of the majority of internet traffic (Munson, 2018) and data generated by volume. This sensor data is typically images, video, or audio, but there are many forms of bespoke sensor data. For example, LIDAR point cloud, RADAR point cloud, and SONAR data is increasingly common in autonomous vehicles (Royo and Ballesta-Garcia, 2019). Infrared data is also increasingly common in satellite data (Tronin *et al.*, 2002). Sensor data is often paired with metadata, such as timestamps or GPS coordinates. In this monograph, we focus on semantic information not present in this metadata. We describe several use cases below.

In the case of text, some text already contains the information needed (e.g., the role of an employee). However, much of the interesting information is not. Consider the example of a court opinion. The judge may use different forms of legal reasoning to justify different parts of the opinion. The type of legal reasoning (and to which paragraphs they apply to) is rarely present as metadata for the court opinion. We focus on this kind of semantic information in this monograph.

## 2.2 Use Cases for Unstructured Data

Many sources of unstructured data are directly related to the "real-world." As such, this data is of great interest to practitioners including business analysts, data scientists, social scientists, legal scholars, urban planners, and others. The types of analyses these practitioners wish to perform are widely varying and have different constraints. We describe several example use cases in detail to motivate the remainder of this monograph.

**Legal scholarship.** Consider the use case of legal scholars attempting to understand how legal reasoning changes over time. In fact, such analyses have been conducted with the aid of AI and ML techniques (Dai *et al.*, 2024; Peters, 2023) to understand the rise of textualism as legal reasoning since the 1980s in America. Such analyses can help frame recent debates around legal practices.

Given the machine-readable text, the legal scholars can then classify what kinds of legal reasoning were applied at the opinion- or paragraph-level. For example, it is widely thought that textualism as a form of legal reasoning has risen and fallen in terms of popularity (Molot, 2006). However, there has not been a large scale, quantitative study of the prevalence of textualism.

Today, performing this kind of analysis at scale would require hundreds of legal scholars to perform manual classification. As we will describe, this kind of analysis is becoming feasible with automatic ML methods.

**Urban planning.** Consider the use case of an urban planner who is studying traffic patterns in a small town. We assume the urban planner has access to traffic camera video footage from around town.

Understanding traffic patterns is a complex topic. The urban planner may be interested in questions ranging from "what is the total number of cars that pass by this intersection per hour" to understand bulk statistics to "show me clips when cars suddenly decelerate in an intersection" for manual analysis.

As we can see, some of these queries are simple, bulk aggregations and others are complex, ad-hoc queries. In fact, even expressing the latter query in a machine understandable format can be challenging.

## 2.3   Machine Learning

In order to answer queries over unstructured data, ML will be a critical tool. We now describe recent advances in ML that enable such queries.

Advances in modern ML methods are driven by deep neural networks (DNNs). DNNs are driven by returns to scale in computation and data. Namely, these DNNs learn from incredibly large amounts of data by performing more and more computation over this data. For example, large language models (LLMs) are powered by a primitive called a *transformer* (Vaswani *et al.*, 2017). The best performing LLM at the time of writing is GPT-4 (Liang *et al.*, 2022). The CEO of OpenAI has publicly stated that GPT-4 costs more than *$100 million* to train (Knight, 2023). Other publicly available LLMs take millions of GPU-hours to train (Touvron *et al.*, 2023). Beyond LLMs, vision models also show returns to data and compute (Zhai *et al.*, 2022).

These ML models can now perform amazing feats. On reasoning benchmarks, state-of-the-art LLMs can outperform human experts (Google, 2023). On vision tasks, including object detection and classification, DNNs can match human performance (He *et al.*, 2017).

Aside from the training costs, these ML models are increasingly more expensive to execute as well (the process of executing an ML model is called "inference"). The most capable models are now proprietary, only accessible behind APIs. As a result, they come at considerable markups in price.

To quantify these costs, we show the cost of using state-of-the-art DNNs to analyze a small town's worth of video (100 cameras for one month) and Wikipedia. In particular, we show the cost of a standard structured query over the same data, the cost of using a self-hosted DNN (using Amazon Web Services, approximated for GPT-4), the cost of using API-gated models, and the cost of using crowd-sourced labor. As we can see, the costs are *up to 10 orders of magnitude higher* compared to standard structured queries. In fact, is it *not possible* to run a model of GPT-4 quality in a self-hosted manner. We show a comparison of these costs in Table 2.1.

**Table 2.1:** Costs of executing queries over unstructured data via self-hosted methods, an ML service, and using human annotators compared to the cost of executing a structured query over similar data. Table taken from Kang (2022).

|                             | Urban planning | Wikipedia     |
| --------------------------- | -------------- | ------------- |
| Structured query            | $0.042         | $0.000026     |
| Self-hosted ML (AWS)        | $380,000       | $59           |
| ML service (GCP, OpenAI)    | $18,000,000    | $300,000      |
| Human annotation (Scale AI) | $630,000,000   | $320,000,000  |

Thus, we can see that the cost of deploying ML inference is infeasible at large scale for most organizations.

## 2.4 Structured Data Management and Unstructured Data

Techniques, systems, and tools for analysis of unstructured data draw heavily from the long history of structured data management. A full survey of structured data management is outside the scope of this monograph, but we highlight two important areas in structured data management: approximate queries and queries with expensive predicates.

## 2.4.1   Traditional Query Systems for Unstructured Data

There is a long line of work on traditional query systems for unstructured data (Flickner *et al.*, 1995; Ogle and Stonebraker, 1995), which largely focuses on simple features (e.g., color) or manual labels (e.g., movie metadata or human labels of objects in video). This work has been foundational to the understanding of how to query unstructured data, but was studied before automatic methods of data extraction were feasible.

## 2.4.2   Queries with Expensive Predicates

Other work has focused on optimizing queries with expensive predicates (Hellerstein and Stonebraker, 1993; Hellerstein, 1998; Joglekar *et al.*, 2015; Kemper *et al.*, 1994). This work largely focuses on optimizing exact query plans to handle massive disparities in processing costs. Much of this work can be directly applied to queries that use ML models as UDFs.

However, as we have shown, exhaustive execution is infeasible for many use cases. As such, most of the work to optimize queries over unstructured data focuses on approximate queries in one form or another.

## 2.4.3   Approximate Queries

One important area of research in standard structured data management is approximate query processing (AQP) (Chaudhuri *et al.*, 2017; Li and Li, 2018). In AQP, a user is interested in answering a query but can tolerate error in the query results. This body of research has focused on aggregate queries.

There is an enormous range of techniques to accelerate approximate queries. All AQP techniques take a sample of the data or a summary of the data and use the sample/summary to produce an approximate answer.

Broadly, these techniques fall under two categories (Li and Li, 2018). The first category is precomputing information to subsequently use for downstream queries. This precomputation can result in synopses (e.g., histograms, wavelets, data cubes, pre-computed samples, or other summaries) or other summary statistics (Agarwal *et al.*, 2013; Acharya *et al.*, 1999; Piatetsky-Shapiro and Connell, 1984; Poosala *et al.*, 1996; Cormode *et al.*, 2009; Guha and Harb, 2005; Garofalakis *et al.*, 2002; Gan *et al.*, 2020). Unfortunately, this form of AQP is broadly not applicable to unstructured data queries as the structured information of interest is not available ahead of time.

The second category of AQP techniques is the various forms of online sampling in which samples are drawn for each new query (Hellerstein *et al.*, 1997). These samples can be drawn in a streaming fashion or in batch. Online sampling is the most relevant part of the structured AQP research. However, as with precomputation, online sampling that uses precomputed sources of information (such as indexes) are not applicable in the unstructured data setting.

Given these techniques to accelerate approximate queries, a critical aspect of usability is the specific semantics of the error on the query results. These semantics range greatly, from best-effort to providing confidence intervals (Park *et al.*, 2018). We will discuss error semantics in greater detail in Section 6.

In the remainder of the monograph, we will turn to specific methods of expressing and answering unstructured data queries.

# 3

# Architecture

In this section, we describe the architecture of a hypothetical end-to-end unstructured data management system. Although no work implements this system end-to-end, we will use this architecture to place the work in this area in context.

We show the end-to-end architecture in Figure 3.1. Much of these components are also present in standard structured data management systems, such as the storage engine and query optimizer. We discuss salient differences below, with further details in later sections.

The first component is some method of specifying queries (Section 4). Many systems implement some form of standard SQL or an extension of SQL to express queries. However, users can also express queries via UDFs that execute ML models or via examples. Expressing queries is especially critical for unstructured data since the queries can be difficult to express in standard SQL or even be ambiguous, unlike for relational data.

Once a query is dispatched to the execution engine, there is a wide range of internals that are required to optimize and execute the query. This includes the query optimizer (Section 6), the storage and index layer (Section 7), and the execution layer (Section 8).

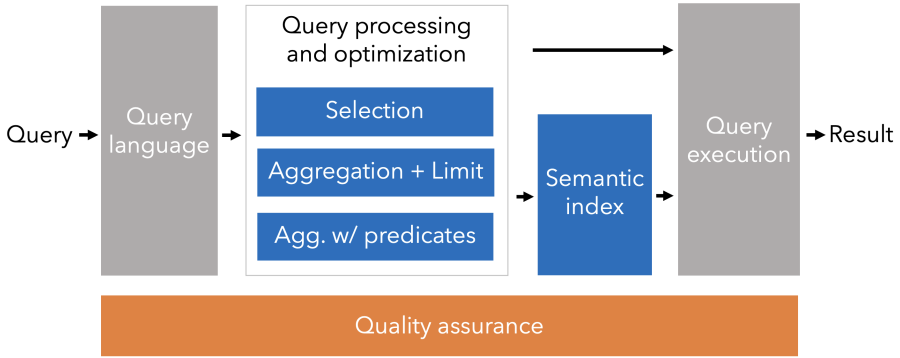

**Figure 3.1:** Overall architecture of a prototypical unstructured data system. Many research projects implement one part of the system or combine aspects.

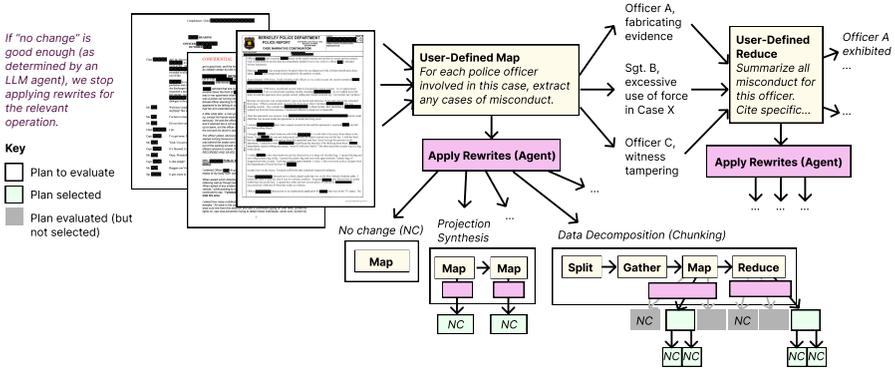

**Figure 3.2:** Architecture diagram of DocETL (Shankar *et al.*, 2024b). DocETL uses LLM agents to rewrite query plans.

Although the components in Figure 3.1 are widely used, newer systems may have different ways of formulating query plans and executing queries. For example, DocETL uses LLMs to rewrite query plans (Shankar *et al.*, 2024b). We show an architecture diagram in Figure 3.2.

Because systems such as DocETL are rapidly being developed, we will primarily focus on the components in Figure 3.1 in this monograph. However, there are many other developing systems and important system components. These components include visualization, which is especially important for video. We now turn to describing the components outlined in Figure 3.1.

# 4

## Expressing Queries

The first step in answering queries over unstructured data is for the user to specify the query. Although seemingly simple, query expression can be challenging for unstructured data. Suppose an urban planner desires to find example video clips of cars suddenly decelerating in an intersection. What does it mean to suddenly decelerate? What does in the intersection mean? In the social sciences, the social scientist may wish to find articles "about" COVID-19. Does this include articles that mention COVID-19 in passing when discussing the economy in 2024?

Unfortunately, mind reading technology is in its infancy, so users must specify the queries directly. In this section, we discuss ways for users to specify complex, unstructured queries.

We begin with assuming that the ML model directly returns the answer of interest. Under this assumption, users can directly use ML models within structured data schemas as UDFs or by mapping columns within a standard relational schema. However, this approach does not handle situations where the ML model does not directly return the information of interest.

To answer more complex queries, researchers have built bespoke techniques. These techniques range from expressing queries by example, directly using natural language, and DSLs for complex video queries.

We will discuss these methods of expressing queries in the remainder of this section.

## 4.1 User-defined Functions

The simplest way to express queries over unstructured data is to assume that there is a programmatic method of extracting the information of interest. This method generally involves calling a machine learning model. For example, an image classifier can return a boolean indicating whether or not a frame of a video contains a car or not.

Given this programmatic method, a user can expose the method as a user-defined function (UDF). These UDFs can be used within traditional data processing engines, such as Postgres or Spark, to answer queries (Armbrust *et al.*, 2015; Crotty *et al.*, 2015). Several ML-focused query engines also expose ML models as UDFs and have wrappers for commonly used APIs, such as OpenAI, HuggingFace, Amazon Rekognition, and more (Xu *et al.*, 2022; MindsDB, n.d.). Research engines have been built based on UDFs as well (Xu *et al.*, 2022; Poms *et al.*, 2018; Kang *et al.*, 2017; Anderson *et al.*, 2019; Lu *et al.*, 2018). Finally, there are commercial products that allow for the expression of ML models as UDFs, including EvaDB and MindsDB.

Many ML-based UDFs are one-to-one mappings between input columns and output columns, such as an image to a boolean (e.g., whether or not the image contains a car or not). These UDFs can be expressed easily. For example, consider extracting sentiment via OpenAI's API. This can be implemented in EvaDB as follows:

```
SELECT ChatGPT(
    "Is the review positive or negative?
    Only reply 'positive' or 'negative'.
    Here are examples.
    The food is very bad: negative.
    The food is very good: positive.",
    review)
FROM postgres_data.review_table;
```

and the output would be as follows:

```
+------------------------------+
|              chatgpt.response |
|------------------------------|
|                      negative |
|                      positive |
|                      negative |
+------------------------------+
```

As we can see, the `ChatGPT` UDF directly maps the review to "positive" or "negative."

Beyond one-to-one mappings, many ML models can produce a variable number of outputs. For example, an object detection DNN can produce zero outputs (if there are no objects in the image) or many outputs (if the scene is busy). Expressing these functions requires returning lists or user-defined table functions (UDFTs). The output of these functions are often more complex to handle within a traditional relational framework. For example, executing an object detection network within EvaDB would look like:

```sql
SELECT id, Yolo(data)
FROM obj
WHERE id < 20
LIMIT 5;
```

and the output would look like

```
+---------+-----------------+-------------------+-------+
| obj.id  |   yolo.labels   |    yolo.bboxes    | yolo.scores |
+---------+-----------------+-------------------+-------+
|   0     | ['car', 'car ... | [[828.7, 277 ... | [0.91, 0.85, ... |
|   1     | ['car', 'car ... | [[832.3, 278 ... | [0.92, 0.85, ... |
|   2     | ['person', ' ... | [[835.7, 279 ... | [0.91, 0.84, ... |
|   3     | ['car', 'car ... | [[839.3, 279 ... | [0.91, 0.84, ... |
|   4     | ['car', 'car ... | [[843.2, 280 ... | [0.9, 0.85,  ... |
+---+-----------------+-------------------+-------+
```

As we can see, writing a query of the form "select cars where the left corner is within a box" requires writing complex SQL to parse the lists across columns (note that the labels also contains other object classes, such as pedestrians).

Nonetheless, there is growing commercial interest in these systems, including EvaDB and MindsDB.

## 4.2 Direct Schemas

Beyond expressing ML models via UDFs, the outputs of ML models can directly be mapped to columns where the rows are materialized on demand. This idea has been applied specifically to video, but also other settings as well (Kang *et al.*, 2019; Xu *et al.*, 2022; Petersohn *et al.*, 2020).

To understand why this can be useful, consider the object detection example above. Instead of returning a list per attribute per frame, we can directly have columns for the labels, box coordinates, and scores. The example above would instead be:

```
+--------------+---------+-------+---+
|   id   |   label   | xmin  | score |
+--------------+---------+-------+---+
|   0    |  'car'    | 828.7 | 0.91  |
|   0    |  'car'    | 755.3 | 0.85  |
|   1    |  'car'    | 832.3 | 0.92  |
|   1    |  'car'    | 756.6 | 0.85  |
|   2    |  'person' | 835.7 | 0.91  |
+---+---------+---------+-------+---+
```

(with the remainder of the bounding box coordinates omitted for brevity). With this schema, selecting cars where the xmin is less than 500 would simply be:

```
SELECT *
WHERE xmin < 500
   AND label = 'car'
FROM obj;
```

The BLAZEIT system introduced a schema specifically for objects in visual data (Kang *et al.*, 2019). It further introduced specialized keywords to aid in computing statistics within a frame.

Beyond video, direct schemas can be used with any form of unstructured data. For text, a business analyst may be interested in extracting topics and sentiments. For audio, a social scientist may be interested in extracting interruptions. Beyond text and audio, semi-structured documents like PDFs are also common.

Although these methods of expressing queries are useful when the user knows exactly how to extract the information, this is not always the case. As such, researchers have developed other methods of allowing users to query for difficult to specify information.

## 4.3   Query by Example

Consider an ecologist who is studying the feeding patterns of various animals. One ecologist may place cameras in the field by a bush to find hummingbirds feeding at the bush (Kang *et al.*, 2021a). Another may place cameras on deer to understand the feeding patterns of deer in the wild (Zhang *et al.*, 2023). We show examples of hummingbirds in Figure 4.1.

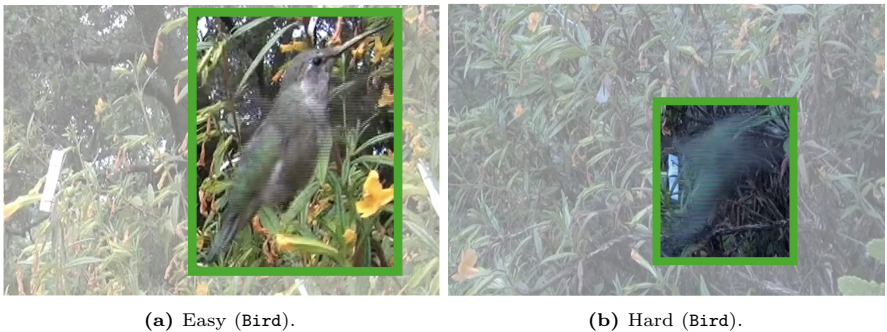

(a) Easy (`Bird`).                    (b) Hard (`Bird`).

**Figure 4.1:** Examples of an easy and hard hummingbird to classify. Taken from Kang *et al.* (2021a).

In both of these examples, the end users (the ecologists) can identify the events of interest but there are often no ML models that can directly identify the events of interest.

In order to find these events of interest, researchers have proposed systems to *query by example* (Zhang *et al.*, 2023; Mell *et al.*, 2021; Kang *et al.*, 2021a). Querying by example requires the end user to interact with the system, either in a batch setting or in an interactive setting. In this setting, the events of interest are typically rare. For example, hummingbirds may be feeding less than 0.1% of the time at a particular bush.

There are many forms of systems that allow users to query by example. One class of systems extends or leverages various forms of *active learning* (Settles, 2009). In active learning, a classifier is iteratively trained to identify a positive class. However, standard active learning methods focus on training a high quality classifier, which may not identify rare events well. As a trivial example, a classification task with 99% negative examples would achieve 99% accuracy by always guessing the negative class.

Since query by example typically focuses on selection queries, we are interested in finding positive examples. As a result, the algorithms that leverage active learning for query by example use structures present in the data for sample efficiency.

One system, EQUI-VOCAL (Zhang *et al.*, 2023), focuses on events over video. EQUI-VOCAL operates over scene graphs in videos, in which relationships between objects within a frame and across frames. We show an example in Figure 4.2. Given these relationships and several positive and negative examples, EQUI-VOCAL will synthesize declarative queries over the schema of relationships that match the positive examples but do not match the negative examples. Because the schema of relationships has so much structure, not many examples are required to synthesize such queries. In simple cases, as few as 12 total examples (2 positive, 10 negative) can be sufficient to synthesize these queries.

Other systems, such as QUIVR (Mell *et al.*, 2021), also synthesize queries. The methods of selecting which examples to ask users to annotate (i.e., active learning) and pruning declarative queries vary from system to system.

Beyond synthesizing declarative queries, other work focuses on training a classifier specifically for selection, and leveraging structure in the data to find more of the positive events (Kang *et al.*, 2021a). For example, hummingbird visits to bushes are temporally correlated. The temporal correlation can be used to find feeding events once a single feeding event is found.

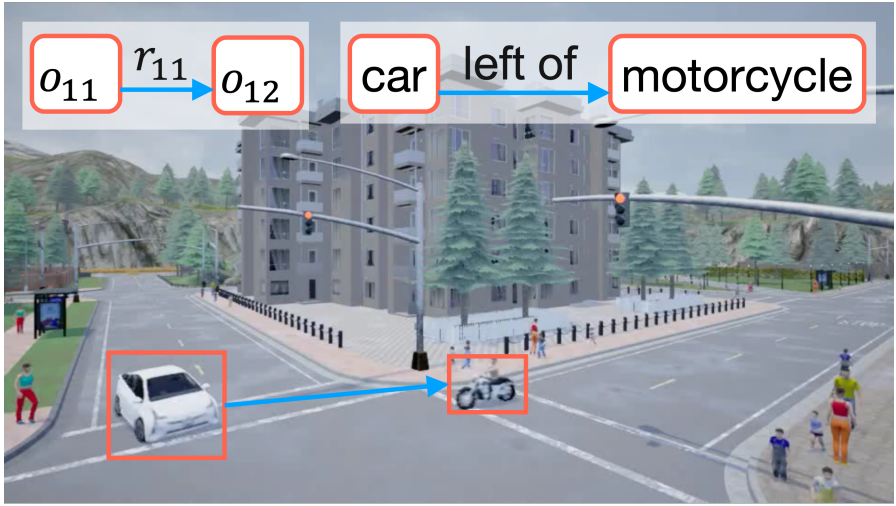

**Figure 4.2:** Example of a region graph in EQUI-VOCAL. A scene graph is not shown. Taken from Zhang *et al.* (2023).

## 4.4 Expressing Queries via Natural Language

Beyond the ways of expressing queries discussed so far, other systems also allow users to express queries via natural language directly (Thorne *et al.*, 2021; Chen *et al.*, 2023; Lei *et al.*, 2021). At a high-level, expression of queries via natural language follows similar patterns to standard SQL: simple queries (e.g., a single table select-project) are easy to express and execute, but more complex queries require much more machinery.

As such, much of the work focuses on how to handle complex queries. They dynamically interpret and decompose complex user queries into simpler sub-queries, using neural models to directly retrieve and reason about relevant information across modalities (e.g., text, video). These systems strategically manage limitations of current LLMs, such as difficulty with aggregation and scaling to large datasets, through techniques like parallel processing of minimal data subsets, adaptive query decomposition with LLMs, and hybrid neural-symbolic execution strategies.

## 4.5 Video Queries

Many systems focus on allowing queries for specific domains. These domains include audio, text, and visual data. We now present a brief survey on methods to query video data specifically. Note that many of these methods of expressing queries fall under other categorizations described above and can potentially be used for other modalities (e.g., images), but we focus on video as a specific case study of a modality.

Querying video data has had a long history in the data management community (Zhang *et al.*, 2023; Chao *et al.*, 2020; Chen *et al.*, 2021; Liu *et al.*, 2019; Chen *et al.*, 2022; Yadav and Curry, 2019; Kuo and Chen, 2000). However, much of this work focuses on manually curated data (such as metadata about actors in a film) or simple features (such as color histograms). Recent ML methods have allowed the extraction of semantic data automatically and, as a result, demand new methods of querying data.

One line of work focuses on queries over "tracks," which are temporally consistent trajectories of objects over video (Bastani *et al.*, 2020; Bastani and Madden, 2022). This line of work can answer a range of queries, including queries about when objects enter and leave a video, when objects suddenly accelerate, and others.

Other work focuses on creating domain-specific languages (DSLs) or purpose-built query languages for video queries (Fu *et al.*, 2019; Yadav and Curry, 2019; Yu *et al.*, 2023; Xiao *et al.*, 2023; Wu *et al.*, 2024a). These DSLs are typically embedded in some other language, such as complex-event processing or Python. One system, Rekall (Fu *et al.*, 2019), is embedded in Python. It defines standard methods of processing object tracks and combining information across object tracks. We show an example in Figure 4.3. Another system, VQPy (Yu *et al.*, 2023), also embeds in Python. VQPy has been productionized at Cisco. We show an example of the syntax in Figure 4.4.

Beyond these systems, there are a wide range of systems for querying semantic information in video (Kuo and Chen, 2000; Chen *et al.*, 2022; Liu *et al.*, 2019; Chen *et al.*, 2021; Chao *et al.*, 2020). We provide a summary of the features of these works in Table 4.1 (which was taken from Zhang *et al.*, 2023).

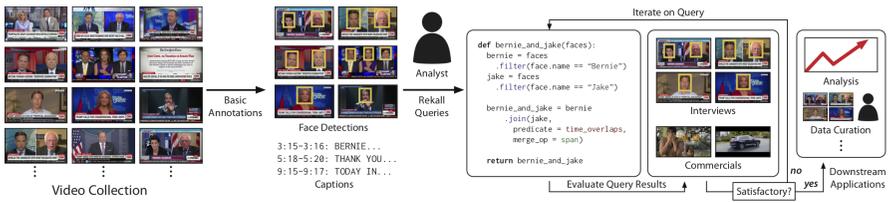

**Figure 4.3:** Example of how to use Rekall. Taken from Fu *et al.*, 2019.

```python
from vqpy.lib.query import CollisionQuery, SpeedQuery
from vqpy.lib.query import SequentialQuery

query HitAndRun(vqpy.Query):
    def __init__():
        self.car = Car()
        self.person = Person()
        car_hit_person = CollisionQuery(
            subqueries = [self.car, self.person]
            dist_threshold = 0.1)
        car_run_away = SpeedQuery(car, velocity_threshold)
        self.sequential = SequentialQuery(
            subqueries=[car_hit_person, car_run_away],
            time_window="10s",
            id_match=(car_hit_person.car, car_run_away.car)
        )

    def video_constraint():
        return self.sequential

    def video_output():
        return self.car.license_plate
```

**Figure 4.4:** Example of how to use VQPy. The query searches for hit-and-run events. Taken from Yu *et al.* (2023).

**Table 4.1:** Comparison between compositional video analytics systems. Taken from Zhang *et al.* (2023).

|  | SVQ++ | Chen *et al.* (2021) | Caesar | STAR | VidCEP | CVQL | Quivr | Rekall | EQUI-VOCAL |
|---|---|---|---|---|---|---|---|---|---|
| Object detection | ✓ | ✓ | ✓ | ✓ | ✓ | ✓ | ✓ | ✓ | ✓ |
| Object tracking |  | ✓ | ✓ | ✓ | ✓ |  | ✓ | ✓ | ✓ |
| Relationship | ✓ |  | ✓ | ✓ | ✓ | ✓ | ✓ | ✓ | ✓ |
| Attribute |  |  |  | ✓ | ✓ |  |  | ✓ | ✓ |
| Conjunction |  | ✓ | ✓ | ✓ | ✓ |  | ✓ | ✓ | ✓ |
| Sequencing |  |  | ✓ | ✓ | ✓ |  | ✓ | ✓ | ✓ |
| Iteration |  | ✓ |  |  |  | ✓ | ✓ | ✓ | ✓ |
| Window |  | ✓ |  |  | ✓ |  |  | ✓ | ✓ |
| Query by example |  |  |  |  |  |  | ✓ |  | ✓ |

Once a query is specified, a data management system must execute such a query efficiently. We now turn to methods of executing such queries efficiently.

# 5

# General Query Optimization

In this section, we discuss general query optimization techniques for unstructured data queries before turning to query optimization techniques specific for approximate queries. Recent advancements in query optimization for unstructured data, particularly video content, have focused on leveraging domain-specific knowledge and declarative interfaces to enable users to aid in query optimization. We highlight three systems in this area: VIVA (Kang *et al.*, 2022b), Relational Hints (Romero *et al.*, 2022), and ClueVQS (Chao *et al.*, 2024).

VIVA and relational hints deploy a simple idea: user-defined declarative specifications of relationships between machine learning models that guide query optimization. The core idea is to enable users to define hints that are obvious with domain knowledge, but are unknowable to the query optimization system. For example, consider analyzing a video of a street corner and looking for a specific license plate number. Empty video frames can be filtered out before an expensive license plate detection method is applied.

These hints can range from substitution hints (e.g., substitute a cheaper model when lower accuracy can be tolerated) to conditional hints (e.g., only invoke a specialized license plate detector if there is a car in the frame of a video). By using these hints, a query optimization system can reorder operations or even skip certain operations entirely for an order of magnitude speedup or more.

ClueVQS extends this idea by introducing clues, a framework where users explicitly declare domain-specific knowledge for the purpose of guiding query optimization. Clues, categorized into types such as MONO-TONIC or DISJOINT. Internally, ClueVQS focuses on optimization strategies to handle issues such as exponentially many query plans. It uses techniques such as pruning to handle such issues. Similar to relational hints, clues can be applied to save orders of magnitude of runtime for complex queries.

We now turn to optimization techniques for approximate queries.

# 6

## Approximate Queries with ML

In order to answer exact queries, any data management system must execute the most accurate model on *all* of the relevant data. This is untenable in many cases, as we have described in Section 2.

Instead, researchers and practitioners have turned to *approximate queries* to reduce the cost of answering queries. Approximate query processing (AQP) has had a long history in the structured data management community (see Section 2). However, unstructured analysis requires rethinking the traditional approximate query stack.

In traditional structured data systems, it is almost always the case that the full query can be run exactly if necessary. Thus, most approximate queries are used primarily for exploratory queries or interactive analysis. Because the full query can be executed, the error guarantees are often not strict. For example, many structured approximate query systems only provide *post-hoc* guarantees on error (Pol and Jermaine, 2005; Galakatos *et al.*, 2017; Kandula *et al.*, 2016). Namely, any error bounds are provided after the query is executed: the user cannot specify an error bound up front. If the user is not satisfied with the error bound, they can simply rerun the full query.

Unfortunately, this is not the case for unstructured data queries. The cost is simply too high to run exact queries in many circumstances. Thus, many unstructured data systems primarily rely on approximations.

In this section, we describe the query semantics behind approximate queries and provide an overview of several methods to accelerate approximate queries tailored to unstructured data systems.

## 6.1 Approximate Query Semantics

One important consideration for approximate queries that is not present in exact queries are *error semantics*. Since approximate queries only return approximations to exact answers, users are often interested in understanding what guarantees are provided with queries.

The semantics for error vary wildly between AQP systems. We show an overview of the error semantics in Table 6.1. As mentioned, few of these systems accept errors, and the ones that do generally pre-compute on the whole dataset to accelerate queries.

**Table 6.1:** Semantics of unstructured data systems.

| System | Best-effort | Outputs errors | Accepts errors | Aggregation | Selection |
|---|---|---|---|---|---|
| NoScope | ✓ | ✗ | ✗ | ✗ | ✓ |
| Probablistic Predicates | ✓ | ✗ | ✗ | ✗ | ✓ |
| Tahoma | ✓ | ✗ | ✗ | ✗ | ✓ |
| SUPG | ✗ | ✓ | ✓ | ✗ | ✓ |
| BlazeIt | ✗ | ✓ | ✓ | ✓ | ✗ |
| MIRIS | ✓ | ✗ | ✗ | ✓ | ✓ |
| ThalamusDB | ✗ | ✓ | ✓ | ✓ | ✗ |

Because pre-computation over the entire dataset is infeasible in many cases, many data systems for unstructured data are approximate *by default*. In fact, due to the cost of ML methods, many of these systems *do not provide any guarantees on accuracy at all*.

For example, systems that accelerate queries over tracks of objects or for detection of objects in video only provide *best-effort* semantics on accuracy (Bastani *et al.*, 2020; Kang *et al.*, 2017; Anderson *et al.*, 2019; Lu *et al.*, 2018; Bastani and Madden, 2022). Namely, *the query answers can be arbitrarily far off from the "ground truth"*. We discuss how these systems attempt to provide heuristics on accuracy below.

Beyond these systems, other work provides guarantees on query accuracy (Kang *et al.*, 2019; Kang *et al.*, 2020; Kang *et al.*, 2021b; Russo *et al.*, 2023; Jo and Trummer, 2024). Much of the work that provides guarantees on accuracy accepts error targets as part of the query semantics. These error targets can be absolute error targets or relative error targets. As an example of the semantics, consider an absolute error target for a query computing the fraction of reviews that are positive. Assuming positive reviews are 1 and negative review are 0 for simplicity, the query might look like:

```
SELECT AVERAGE(sentiment)
FROM reviews
ERROR WITHIN 0.02
CONFIDENCE 95%;
```

As we can see, there are two new keywords: `ERROR WITHIN` and `CONFIDENCE`. Because the error target here is *absolute*, this would translate to the query being answered within 2% of the true answer. Importantly, this is not a relative error: if the true answer was 1%, a relative error would need to be within 0.02%. Nonetheless, this form of error semantic is useful for exploratory queries.

Other systems provide semantics for relative error (Jin *et al.*, 2024). These error semantics are more standard. The same query above with a 5% relative error target may look like:

```
SELECT AVERAGE(sentiment)
FROM reviews
ERROR PERCENT 5%
CONFIDENCE 95%;
```

Beyond traditional aggregate queries, several systems optimize queries for approximate *selection*. In this setting, the user issues a *selection* query with a predicate, which is often highly selective. Instead of returning an aggregate statistic, the result of these queries are a set of tuples. Because the result is a *set of tuples* instead of an aggregated statistic, the errors semantics must necessarily be different.

The simplest error statistic for selection is overall accuracy. Across the entire table, a record is assigned a 1 if it is correctly included or excluded from returned set and a 0 otherwise. The accuracy is the average value across the 0/1 values per record. Given this error statistic, the semantics can be best effort (i.e., no guarantees), in expectation, or with a given failure probability.

There are two more common error statistics, often used in conjunction. The first is the *recall* of the returned set of tuples. The recall is defined to be the fraction of records that match the predicate that are in the returned set. A recall of 1 means that all records matching the predicate are in the returned set. A recall of 0 means that none of the records are in the returned set. The second is the *precision* of the returned set of tuples, which is the fraction of records in the returned set that match the predicate.

When used in isolation, these error statistics have limited use. Any system can trivially achieve a perfect recall or precision with unlimited budget by exhaustively executing the ML model. Alternatively, the system can return all the records (for perfect recall but poor precision) or return no records (for perfect precision by poor recall).

As a result, it is common to combine a recall or precision target with some other metric. Common combinations include a recall target with a minimum precision, a recall target with a fixed budget (with the goal of maximizing precision), a precision target with a minimum recall, and a precision target with a fixed budget (with the goal of maximizing recall) (Kang *et al.*, 2020). These error statistics can also be combined heuristically (i.e., without any guarantees), in expectation, or with a given failure probability.

Finally, there are bespoke systems that define their own error semantics, either for the types of queries mentioned above or for new kinds of queries (Russo *et al.*, 2023).

We now turn to methods of accelerating these approximate queries.

## 6.2 Proxy Models

The first class of methods we will discuss leverage *proxy models*, which are approximations to expensive "ground truth" or "oracle" models. We provide an in-depth discussion of how to construct proxy models in Section 7 and simply discuss the interface of proxy models and using them for downstream queries.

There are generally two classes of proxy models: proxies that compute the same outputs as the oracle model and proxies that compute some fixed function of the oracle model outputs.

Consider the object detection example. A proxy that computes the same output as the oracle object detection model is simply a cheaper object detection model. A proxy that computes a fixed function of the object detection model's outputs may compute the binary label of whether or not a car is in a frame of a video. These forms of proxies are useful when not all of the oracle model's outputs are needed to answer a query: selecting frames when cars are present does not require knowing the objects' positions.

In order for proxy models to be useful, they must be cheaper than the oracle model. This is not difficult to achieve, especially in the second case: classification models can be thousands of times cheaper than object detection models (Kang *et al.*, 2017; Anderson *et al.*, 2019; Lu *et al.*, 2018). Section 7 discusses how to construct these proxy models.

## 6.3 Approximate Aggregation Queries

The first class of queries we will discuss are approximate aggregation queries. In these queries, the user is interested in a summary statistic, such as the average number of cars per frame of a video. In this work, we will focus on *linear* summary statistics, which includes SUM, COUNT, and AVG.

Given a proxy model for the summary statistic, the simplest method of answering the query is to use the proxy directly. For example, if the user issues a query to count the average number of cars per frame of video, we can simply use the proxy to directly compute the answer (Kang *et al.*, 2019). Unfortunately, this method of answering queries gives *no* guarantees on query accuracy. In fact, using proxies directly can give arbitrarily bad query results (Kang *et al.*, 2020), which is unacceptable in many cases.

The simplest method of achieving error semantics is to uniformly sample from the unstructured data, execute the oracle method, and return the value on the sample (adjusted for sampling). The error semantics can be achieved using standard confidence interval computations. Although functional, uniform sampling can be inefficient.

Instead, we can combine sampling and proxy models by using the proxy models to guide where to sample from the oracle method or to use the proxy model to achieve better estimates (Kang *et al.*, 2019; Kang *et al.*, 2021b). The specific method of using the proxy model depends on the query type.

### 6.3.1 Whole Table Aggregation

For a whole table aggregation, one method of leveraging the proxy is to use it as a *control variate* (Kang *et al.*, 2019; Nelson, 1990). A control variate is a standard tool from traditional statistics.[1] Suppose we wish to compute

$$\mathbb{E}\left[m\right] = \mu$$

where $\mu$ is unknown. In our setting, $m$ is some statistic computed from the output of the oracle model (e.g., the number of cars per frame of video). Generically, suppose we could compute another statistic

$$\mathbb{E}\left[t\right] = \tau$$

where $\tau$ is known. In our setting, $t$ is the proxy model and $\tau$ is the estimated statistic from the proxy model. Then,

$$m^{\star} = m + c\left(t - \tau\right)$$

is an unbiased estimator for $m$ for any $c$. Using the specific choice of $c$ as

$$c^{\star} = -\frac{\text{Cov}\left(m, t\right)}{\text{Var}\left(t\right)}$$

we have that

$$\text{Var}\left(m^{\star}\right) = \text{Var}\left(m\right) - \frac{\left[\text{Cov}\left(m, t\right)\right]^2}{\text{Var}\left(t\right)} \tag{6.1}$$

$$= \left(1 - \rho_{m,t}^2\right)\text{Var}\left(m\right) \tag{6.2}$$

---

[1]The following presentation is largely adapted from Wikipedia (2024).

where
$$\rho_{m,t} = \mathrm{Corr}\,(m, t)\,.$$

Namely, as the correlation between the proxy and oracle increases, the variance of our estimator decreases.

We show a diagram of a proxy as a control variate and the ground truth in Figure 6.1a. As shown, the proxy may not track the ground truth. Nonetheless, using the proxy is better than not as long as the correlation with the ground truth is high.

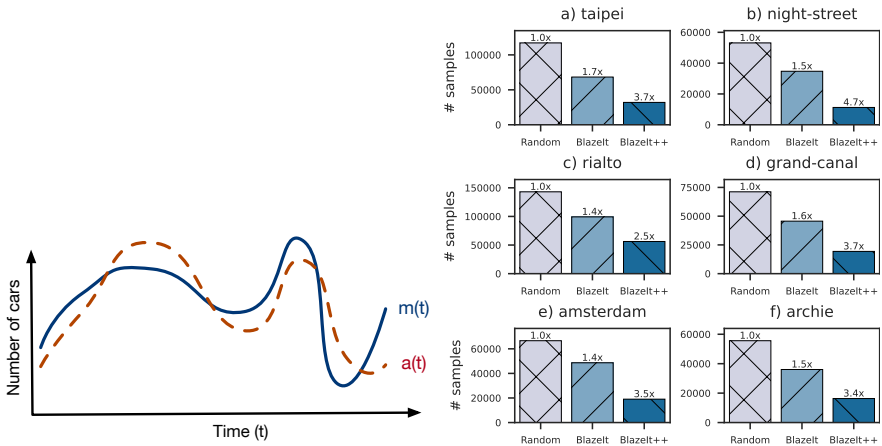

**(a)** Diagram of a ground truth signal ($m(t)$) and a control variate ($a(t)$). Even though $a(t)$ does not perfectly track $m(t)$, it is highly correlated with $m(t)$. Taken from Kang *et al.* (2019).

**(b)** Random sampling vs. two methods of using control variates. Control variates can reduce the number of samples needed for a fixed error by up to 5×. Taken from Kang *et al.* (2024).

**Figure 6.1:** Example of control variates.

Prior work has shown that proxies can be extremely correlated with the oracle model, resulting in up to 5× performance improvements compared to random sampling (Kang *et al.*, 2019; Kang *et al.*, 2024). Furthermore, compared to exact queries, a system like BLAZEIT can be several thousand times cheaper (similar to how standard AQP systems can be orders of magnitude faster than exact queries on structured data).

We show the results of an experiment conducted by Kang *et al.* (2024) on counting the number of cars per frame of a video in Figure 6.1b. As shown, using control variates (BLAZEIT ++) can improve over random sampling by nearly 5×.

However, to apply control variates in an unbiased manner, we must know $\tau$. When aggregating across the whole dataset, this is trivial: we can simply compute the proxy score on every record as the proxy is cheap to compute. However, when computing an aggregation query with an expensive predicate, we cannot compute $\tau$ exactly without first computing the predicate, which prevents us from using control variates in an unbiased manner. We now turn to optimizing aggregation queries with predicates.

### 6.3.2 Aggregation with Predicates

Unfortunately, if we wish to compute an aggregation query *with a predicate computed via an ML model*, we cannot directly apply control variates. In order to use control variates, we must know $\tau$ exactly. However, since the results of the predicate are not materialized, we cannot compute $\tau$. Instead, we can use a proxy model to guide where to sample.

To first understand the intuition, consider standard stratified sampling (Neyman, 1992). Stratified sampling splits the table (alternatively referred to as dataset or population) into discrete, non-overlapping strata. The perfect stratification would have *zero* variance within each strata, which would mean each strata has a constant value. In this case, we could simply sample a single item per strata and use the strata size to compute the statistic of interest. However, if strata have non-zero variance, the optimal allocation will allocate more sampling budget to the strata with higher variance (adjusted for strata size). This is because we have more uncertainty in the strata with higher variance.

In the setting of an expensive predicate, we can similarly stratify the table. However, we do not even know the number of the relevant records per strata. To solve this issue, we can first perform a *pilot sample* that is used to estimate the fraction of records that match the predicate and the variance of the statistic. Given our estimates, we can then estimate the optimal allocation and sample according to the optimal allocation.

Because the sampling is uniform within each strata, and the samples that do not match the predicate are rejected, the estimator is asymptotically unbiased. However, because the allocation of the second stage depends on the random pilot sample, it is difficult to compute confidence intervals.

To compute confidence intervals, we can turn to the bootstrap. However, it is not immediately apparent how or where to apply the bootstrap. For example, should we apply the bootstrap to only the second stage? Only the first stage? If we sample across both, should we change the allocation of the second stage? Answering these questions requires a careful analysis of the sampling algorithm. For example, only applying the bootstrap to the second stage will result in invalid confidence intervals, since it does not take into account the variance from the stochasticity in determining the allocation.

The correct approach requires bootstrapping across both the first and second stage, repeating the full procedure each time. Proving the validity of the bootstrap requires a careful analysis of how the first stage influences the second stage. The technical condition required is Hadamard differentiability of the cumulative distribution function, which Kang *et al.* (2021b) show.

These techniques (two-stage sampling and boostrapping) provide fast query results and valid confidence intervals (Kang *et al.*, 2021b). Compared to uniform sampling, a system like ABae can achieve up to 2× cheaper cost.

We show an example evaluation of ABae in Figure 6.2 on a variety of datasets and input modalities, including text and images. The figure compares uniform rejection sampling with ABae on the metric of root mean-squared error (RMSE). As shown, ABae can outperform uniform rejection sampling by up to 1.5× on RMSE at the same budget, or achieve the same RMSE with 2× fewer samples.

### 6.3.3 Discussion

As we have seen, sampling combined with proxy models can accelerate various kinds of aggregation queries. However, different query types require different forms of sampling algorithms! This trend also applies to selection queries and streaming queries as we discuss below.

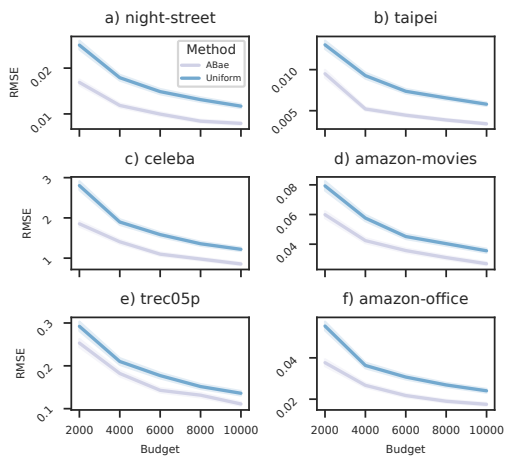

**Figure 6.2:** RMSE of uniform rejection sampling and ABAE. ABAE can outperform uniform rejection sampling by up to 1.5× on RMSE at the same budget, or achieve the same RMSE with 2× fewer samples. Taken from Kang *et al.* (2021b).

## 6.4   Approximate Selection Queries

Beyond aggregations, users are also interested in selection queries with predicates. Selection queries are typically used to find rare events for manual analysis when applied to large unstructured datasets. For example, ecologists at Stanford University are interested in finding hummingbirds visits to a bush in a large quantity of field video (>10TB), where hummingbird visits are <0.1% of the video (Kang *et al.*, 2021a).

Selection queries over rare events are challenging for sampling algorithms. Unlike aggregations, the result from a selection query is a set of tuples. Thus, uniform sampling by itself cannot guarantee any semantics on recall or precision.

As before, we will leverage proxy models to accelerate selection queries. Throughout this section, we will assume that a proxy model returns a continuous value between 0 and 1 per record (i.e., proxy score) indicating how likely the proxy believes the record matches the predicate.

We will first describe how to optimize selection queries with best-effort error semantics before describing how to accelerate selection queries with guarantees on recall and precision.

### 6.4.1   Approximate Selection with Best-effort Semantics

Suppose we had a proxy that returned exactly the result of the expensive predicate. Then, we could simply use the proxy score to return all records that match the predicate. However, exact proxies essentially never occur in practice.

Nonetheless, the example above shows a key desiderata of proxy models: they should closely match the oracle model. There are many ways to quantify "closeness" of the proxy to the oracle, but two useful measures are *calibration* and *sharpness*. A perfectly calibrated proxy is one where the proxy reflects the true probability that the predicate matches and calibration measures the discrepancy from perfect calibration. Sharpness measures how close the values are to 0 and 1. Thus, a perfectly calibrated and perfectly sharp proxy would exactly return the result of the expensive predicate.

Given a proxy, the simplest algorithm to perform selection is to order the records by proxy score, pick some cutoff and return all records above the cutoff (Kang *et al.*, 2017; Anderson *et al.*, 2019; Lu *et al.*, 2018). In this scheme, the complexity is entirely in picking the cutoff. A heuristic method of picking a cutoff involves using a validation set, but this does not provide any guarantees on the recall or precision of the returned set.

The naive method can be extended by choosing two cutoffs: a lower and upper cutoff. Records above the upper cutoff are directly returned. The oracle is queried on records between the lower and upper cutoff. Similarly, the complexity lies in choosing the cutoffs.

There are many methods to choose cutoffs and which proxy models to apply (Kang *et al.*, 2017; Anderson *et al.*, 2019; Lu *et al.*, 2018). These methods range from cost-based optimization to using validation sets.

One simple method of choosing a cutoff is to select some data as a validation set. The oracle is exhaustively executed on the validation set and the empirical cutoff is chosen from the proxy scores on the validation set. We show an architecture diagram of an end-to-end process of using a validation set in Figure 6.3. This system, NOSCOPE, uses the validation set to determine which records query the oracle on.

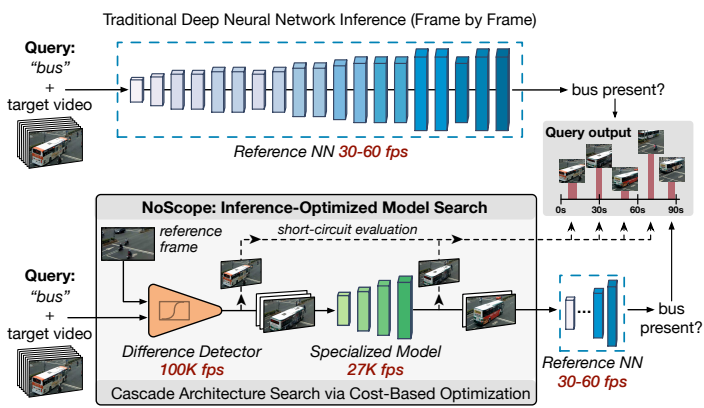

**Figure 6.3:** Architecture diagram for NoScope, a system that uses a validation set to select thresholds for the proxies. NoScope sends records it is unsure of to the oracle for confirmation. Taken from Kang *et al.* (2017).

Using a validation set can result in enormous speedups. For example, prior work can accelerate queries by 10-3,000× compared to exhaustively using the oracle (Kang *et al.*, 2017; Anderson *et al.*, 2019; Lu *et al.*, 2018).

However, the bulk of these methods are heuristic and do not provide guarantees on recall or precision. These guarantees are critical for scientific applications and mission-critical business decisions. As such, we now turn to methods of approximate selection queries with guarantees.

### 6.4.2 Approximate Selection with Guarantees

To understand how an *approximate* selection query can provide guarantees on recall, we first reiterate the semantics of a guarantee on recall. For a recall target of 90%, the goal is to return a set of records such that the returned set contains at least 90% of the records that match the predicate. Importantly, there are no semantics on guarantees on precision.

As such, an algorithm that always returns the entire table will always achieve a recall of 100%. However, this algorithm has poor precision. Thus, for a recall target, we aim to maximize precision.

**Achieving recall targets.** As before, we will order records by proxy score, choose a cutoff, and return all records above a cutoff. Instead of using a validation set, which does not provide guarantees on recall, we will instead using a sampling scheme to select a cutoff that guarantees a recall.

The intuition behind the sampling strategy is as follows. First, suppose we know exactly the number of records matching the predicate (say $M$ records). Then, we simply need to choose a cutoff such that the number of matching records above the cutoff is at least $0.9M$.

To do so, we first consider uniform sampling. We can uniformly sample from the dataset, which will return a set of records. As we described above, one method that achieves *average* recall is to simply take the empirical cutoff of the proxy score that achieves the desired recall in this sample and return all records above that cutoff. However, it does not guarantee a failure probability on recall.

Instead, we can form a confidence interval from the *positive* records. The intuition is as follows. First, we take the empirical cutoff achieving the desired recall (call it $\tau$) as before. We can compute a lower bound on the number of positive records above the cutoff, an upper bound on the number of positive records below the cutoff, and use that to compute a corrected $\tau'$. Kang *et al.* (2020) present a complete algorithm and validity justification.

The primary choices for this general method are the choice of sampling method, proxy and confidence interval computation method. Other work has explored different forms of computing valid confidence intervals (Pol and Jermaine, 2005).

To sample more efficiently, recent work has explored importance sampling (Kang *et al.*, 2020). Standard importance sampling uses weights proportional to the proxy score. Intuitively, this sampling scheme samples more from where the proxy score is closer to 1 and less when it is closer to 0. However, if we assume a sharp and calibrated proxy, the optimal weights are actually proportional to the *square root* of the proxy score (Kang *et al.*, 2020). This can result in substantially higher precision at a fixed recall target, up to a $20\times$ improvement. We show the recall at various precision targets for SUPG in Figure 6.4, taken from Kang *et al.* (2020).

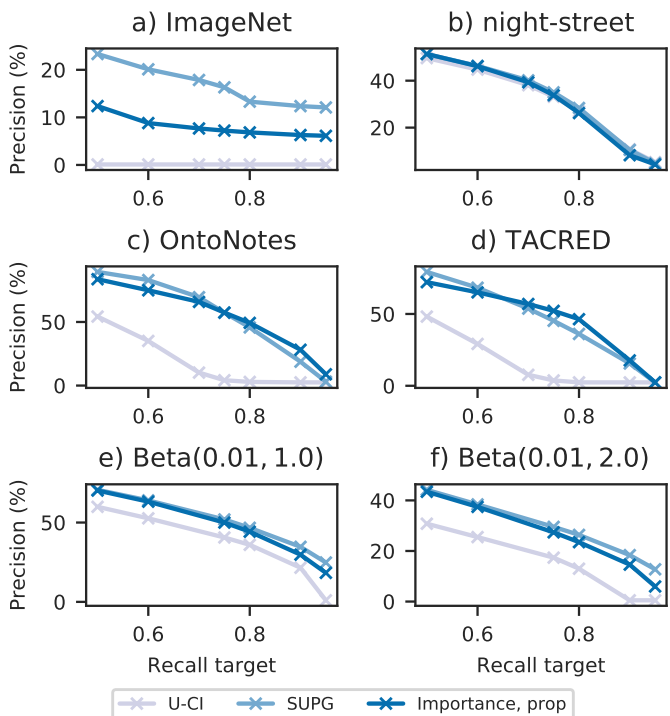

**Figure 6.4:** Precision at various recall targets for SUPG on six datasets. SUPG outperform uniform sampling and standard importance sampling. Taken from Kang *et al.* (2020).

We show a diagram of the intuition behind the algorithm in Figure 6.5. As shown, using uniform sampling without corrections results in invalid results (i.e., results that do not contain enough records) and inefficient sampling.

**Achieving precision targets.** Although less common, users may also be interested in achieving a precision target while maximizing the recall (e.g., return a set of records that are at least 95% positive). From a theoretical perspective, the algorithm is very similar. There are only two substantive differences. First, we compute the precision instead of the recall. Second, we can sweep across the sample and compute multiple cutoffs and choose the one that achieves the highest recall while satisfying the precision. We must correct for the multiple hypothesis testing in the case of testing multiple cutoffs.

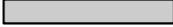

**Figure 6.5:** Diagram of approximate selection with and without guarantees. The naive method uses uniform sampling and the empirical cutoff, which is inefficient and does not produce valid CIs. SUPG instead uses importance sampling and a correction method, achieving efficiency and valid CIs. Taken from Kang *et al.* (2020).

Similar to recall targets, SUPG substantially outperforms baselines on precision targets as well. We show the recall at various precision targets for SUPG on six datasets in Figure 6.6, taken from Kang *et al.* (2020). As shown, the recalls are substantially higher for SUPG compared to uniform sampling.

**Achieving a joint recall and precision target.** Finally, users may be interested in achieving both a recall and a precision target. One method of achieving both targets is to first execute the recall target and then exhaustively execute the oracle method on the records to filter out the ones that do not match the predicate. It is an open problem to determine if there are more efficient methods of achieving both simultaneously.

## 6.5 Approximate Streaming Queries

Beyond batch queries, users may also be interested in streaming aggregates (Jiang *et al.*, 2018; Russo *et al.*, 2023). These queries are common in structured data systems and can be used to find changes in underlying data patterns quickly.

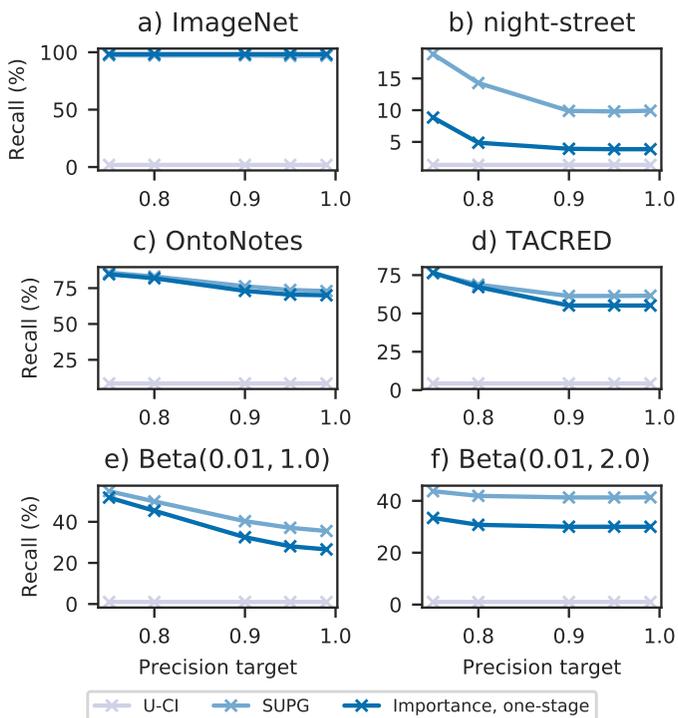

**Figure 6.6:** Recall at various precision targets for SUPG on six datasets. SUPG outperform uniform sampling. Taken from Kang *et al.* (2020).

One simple way to answer a streaming query is to perform reservoir sampling, in which a sample is kept and updated as the stream comes in (Vitter, 1985). When carefully selecting the sample, this can result in a uniform sample, which will give unbiased results.

As with standard uniform sampling in the batch setting, it is often more efficient to use proxies. However, unlike in the batch setting, the data distribution can shift over time. As long as shifts do not occur too quickly, we can use standard techniques to adapt the proxy to the stream. Intuitively, an algorithm like an exponentially decaying weighting will weigh the recent data more. In fact, a similar algorithm can be proven to be efficient if the data distribution does not change too fast (Russo *et al.*, 2023).

## 6.6   Other Approximate Queries

Beyond the specific approximate query processing algorithms discussed
here, recent work has introduced new forms of semantics and algorithms
(Jo and Trummer, 2024; Lab, 2025). For example, ThalamusDB uses a
multi-objective optimizer that considers model costs, data selectivity,
and user preferences to generate optimal execution plans, using zero-
shot classifiers (like CLIP and Sentence-BERT) to prioritized subsets
of data (Jo and Trummer, 2024).

Due to the cost of AI queries, approximate query processing is an
active area of research and we anticipate that these techniques will
evolve over time.

# 7

---

## Proxies, Indexes, and Storage

---

As we have described in the previous section, proxies can be used to dramatically accelerate unstructured data queries. Furthermore, the quality of a proxy can often be precisely measured as it relates to query runtime or quality. Thus, automatically creating high quality proxies is critical to high performance queries.

In this section, we describe methods of generating high quality proxies and storing data for fast queries. We first discuss methods of generating proxies on a per-query basis. While effective, per-query proxies add the training cost on a per-query basis. To reduce these costs, recent work has developed semantic indexes to accelerate many kinds of queries. Aside from semantic indexes, there are other forms of indexes used for modern unstructured data processing, which we discuss briefly. Finally, we discuss storage for unstructured data queries.

### 7.1 Proxies from Scratch

As described in the previous section, different queries work best with different kinds of proxies. For example, selection queries expect a score from 0 to 1 and aggregation queries expect an approximation to the statistic (i.e., a number). Given these different demands, the simplest method of creating proxies for different queries is to generate a proxy per query.

Much of the work in the data management community for creating proxies from scratch focuses on *model specialization*. Model specialization creates a substantially cheaper model for the particular task and data distribution at hand. For example, the task of classification (e.g., whether or not a car is in a frame of video) is simpler than the task of object detection (e.g., determining all of the cars and their positions).

One of the simplest methods of performing model specialization is to take a sample from the target dataset, label this sample with the oracle method, and train a smaller model based on the sample (Kang *et al.*, 2017; Anderson *et al.*, 2019; Lu *et al.*, 2018; Kang *et al.*, 2019). There are different methods of choosing the sample from the target dataset. Furthermore, the sample can be generated at query time or at data ingest time.

In addition to training a single proxy model, many methods train multiple proxy models (Kang *et al.*, 2017; Anderson *et al.*, 2019; Lu *et al.*, 2018). In this case, the system must determine which proxy should be used. The methods for selecting the proxy generally involve a combination of quality estimation (e.g., selectivity or correlation to the oracle) and cost estimation (e.g., the throughput of the proxy model compared to the speed improvement of disregarding the oracle).

While fruitful, training proxies from scratch can add unnecessary overheads per query. To reduce this overhead, other work has explored *semantic* indexes.

## 7.2  Semantic Indexes

Indexes are a long studied and deployed technique in traditional databases (Garcia-Molina, 2008). By performing computation ahead of time, we improve query performance, especially if the index is aware of the workload (Wu *et al.*, 2024b; Zhou *et al.*, 2020).

However, traditional structured data indexes are unsuited for semantic queries over unstructured data because the desired information is not present at indexing time. For example, an inverted index can efficiently count the number of specific instances of a word but cannot tell the sentiment of a Tweet. How can we construct indexes for semantic data?

A now widely used technique is to use *embeddings* (i.e., vectors) as a semantic index. Intuitively, a good embedding "contains" information about the underlying unstructured data. Concretely, consider two images which contain a single car going through an intersection at different points in time. Most queries about the semantic contents of the image (e.g., the car locations) will return similar answers on both images. In most circumstances, high quality embeddings would place these two images close together.

Although useful, these embeddings cannot directly answer queries, except retrieval queries. Thus, an important question is how to use them to accelerate queries.

To accelerate queries, several systems have been developed that essentially group similar records together (as measured by the embedding distance) and propagate the results from known records to unknown records (Kang *et al.*, 2022a; He *et al.*, 2020). We show an overview of TASTI, a particular method of generating embeddings, in Figure 7.1.

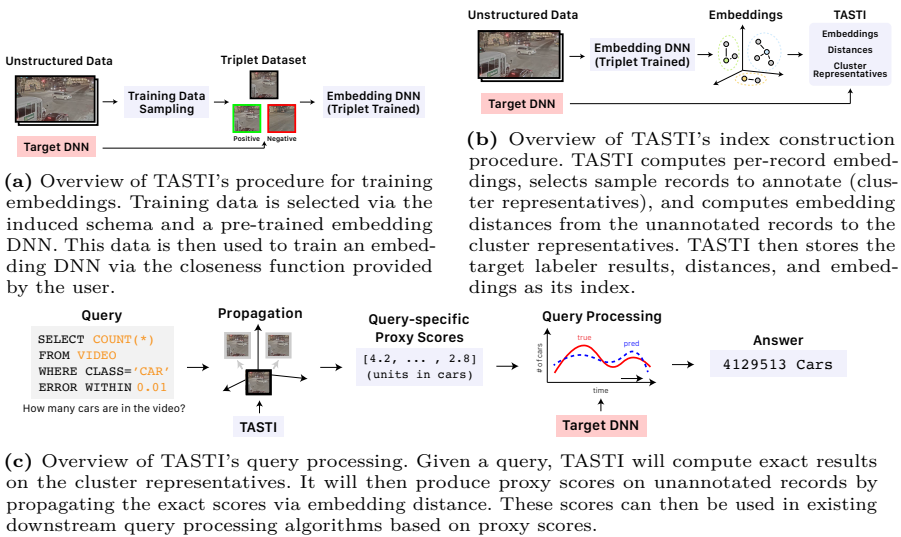

**(a)** Overview of TASTI's procedure for training embeddings. Training data is selected via the induced schema and a pre-trained embedding DNN. This data is then used to train an embedding DNN via the closeness function provided by the user.

**(b)** Overview of TASTI's index construction procedure. TASTI computes per-record embeddings, selects sample records to annotate (cluster representatives), and computes embedding distances from the unannotated records to the cluster representatives. TASTI then stores the target labeler results, distances, and embeddings as its index.

**(c)** Overview of TASTI's query processing. Given a query, TASTI will compute exact results on the cluster representatives. It will then produce proxy scores on unannotated records by propagating the exact scores via embedding distance. These scores can then be used in existing downstream query processing algorithms based on proxy scores.

**Figure 7.1:** TASTI system overview. Taken from Kang *et al.* (2022a).

Deciding how to group records is a key consideration when using semantic indexes. This grouping can be "soft" or "hard."

For example, voodoo indexes are used to accelerate filters (He *et al.*, 2020). In order to construct the index, voodoo indexes use a hierarchical clustering method to form groups. At query time, the optimizer determines how the groups should correspond to passing the filter or not. Since this is a binary decision, the groups are given a label based on runtime information.

Another method, TASTI, focuses on creating an index to generate proxy scores efficiently at runtime. In order to construct the index, TASTI selects a sample of data to label (in addition to generating the embeddings) (Kang *et al.*, 2022a). These labeled data are called cluster representatives. TASTI then computes the distances from the unlabeled records to the cluster representatives and keeps the closest $k$ distances for each unlabeled record. The distances and cluster representatives form the index.

At query time, TASTI generates proxy scores per record by propagating the known results from the cluster representatives to the unlabeled records. Its propagation method is simply an average weighted by the distance.

Surprisingly, high quality embeddings can result in substantially higher quality proxy scores when using TASTI and more accurate results when using voodoo indexes compared to query-specific proxy models. Using high-quality embeddings can also simultaneously be cheaper to construct than using query-specific proxy models. Generally, this is because the indexing model can be more powerful compared to a query-specific proxy as its cost is amortized over many queries.

We show the breakdown of index construction costs for TASTI on a video dataset in Figure 7.2 and its performance on aggregation queries compared to query-specific proxy models in Figure 7.3 (both taken from Kang *et al.*, 2022a). As shown, TASTI is simultaneously cheaper to construct and higher performance at query time.

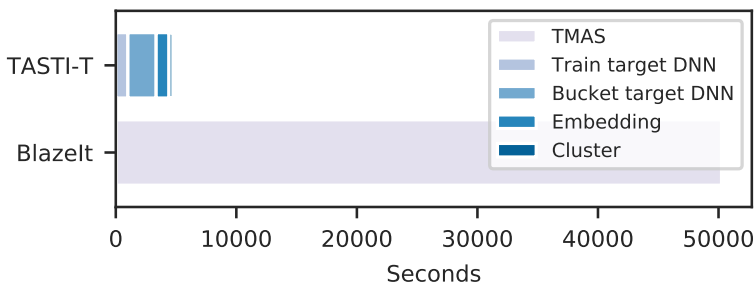

**Figure 7.2:** Cost of constructing a TASTI index compared to the labels for query-specific proxy models. TASTI is cheaper to construct.

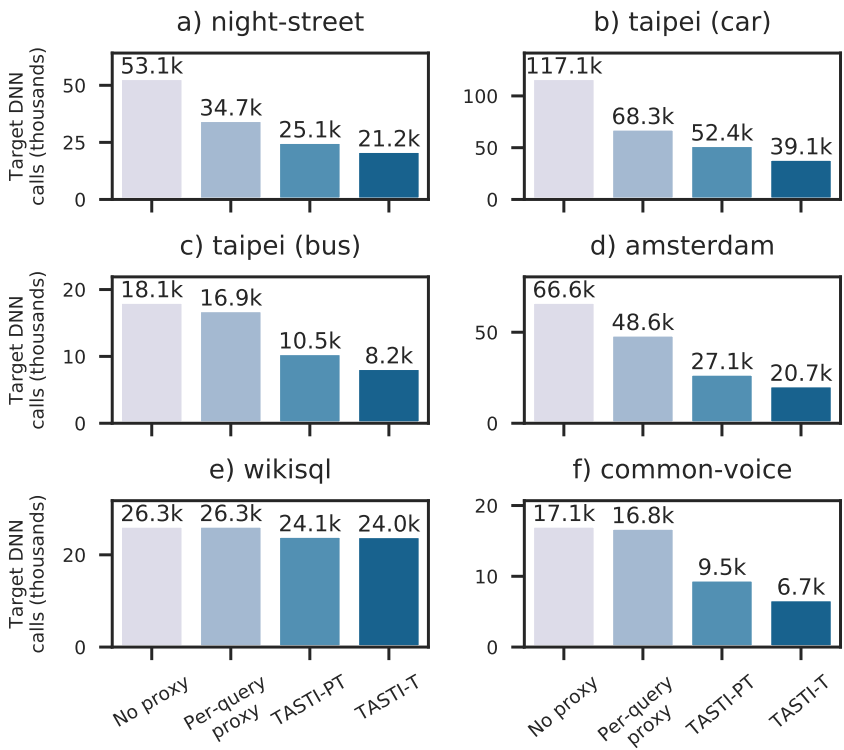

**Figure 7.3:** Performance of TASTI-based proxy scores vs. query-specific proxy models on aggregation tasks on six datasets. As shown, TASTI can outperform by 3× by producing higher quality proxy scores.

## 7.3   Other Indexes and Preprocessing

Aside from semantic indexes, there is a wide range of work that uses various forms of preprocessing to accelerate queries at query time. This work ranges from creating graphs from objects to accelerate spatial relationship queries (Chen *et al.*, 2022), traditional text inverted indexing (Stonebraker and Pavlo, 2024), and others (Hu *et al.*, 2022; Xu *et al.*, 2024).

For example, recent work uses "common sense knowledge" to construct efficient indexes for LIMIT queries (He *et al.*, 2024). Consider searching for a tennis ball in a corpora of video. Tennis balls are more likely to be in videos related to tennis (e.g., on a tennis court) than unrelated videos, such as a video in a lecture hall. One way to operationalize this intuition is to categorize the videos and use a knowledge base to determine which videos to search over first. We show an architecture diagram of Paine, which implements this architecture, in Figure 7.4.

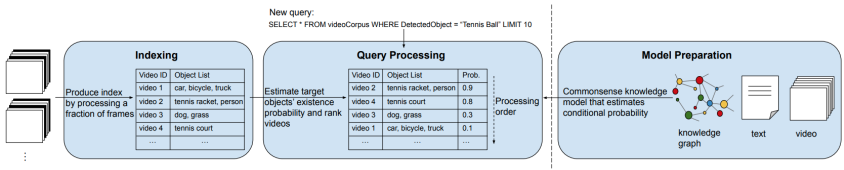

**Figure 7.4:** Architecture diagram of Paine, taken from He *et al.* (2024).

Beyond using specific information about the query structure, a large body of work uses materialized views to reuse intermediate results. We have discussed a number of these systems already (Xu *et al.*, 2022; Jin *et al.*, 2024; Kang *et al.*, 2022b), and such techniques have been long used in traditional query processing with structured data.

Especially as AI techniques have moved towards large, foundation models, we anticipate indexes techniques will quickly evolve.

## 7.4   Storage

Finally, we discuss storage. Storage for modalities such as text is often not a major concern as human-generated text is relatively low volume.

However, automatically generated sensor data can be high volume and thus must more expensive to store. In this section, we will focus on storing video data, but there are also challenges in storing other forms of sensor data, such as LIDAR or satellite data.

Similar to structured data, video is highly redundant. Some of this redundancy is accounted for by using standard video storage such as the H.264 or H.265 encoding formats. However, these video formats focus on high visual fidelity, which is often not required for visual analytics.

Consider Figure 7.5. Although Figure 7.5b has substantially lower resolution, it is still easy to determine that there is a car in the image. Furthermore, much of the image is irrelevant to analytical queries about objects in the image (as highlighted).

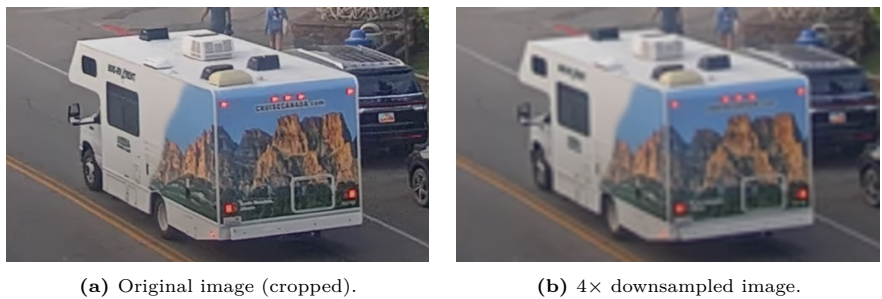

**(a)** Original image (cropped).    **(b)** 4× downsampled image.

**Figure 7.5:** Original vs. 4× downsampled image (one side). Although blurry, it is easy to determine there is a truck in the downsampled image.

Several systems focus on leveraging this redundancy to reduce the storage requirements or speed up data loading from disk (Daum *et al.*, 2021; Xu *et al.*, 2019; Haynes *et al.*, 2021). The intuition behind these systems is to trade off visual fidelity while maintaining high query accuracy by exploiting redundancy and removing unnecessary parts of the video. We show the system diagram of TASM, one such system that leverages redundancy in Figure 7.6 (taken from Daum *et al.*, 2021). These systems can reduce query costs, as we show in Figure 7.7 (taken from Daum *et al.*, 2021).

The redundancy or selected areas in the video can come from overlapping camera views (Haynes *et al.*, 2021) or focus areas where motion is happening (Daum *et al.*, 2021). Typically, these systems update the storage and cache as the query workloads evolve and have some method of efficiently ingesting video.

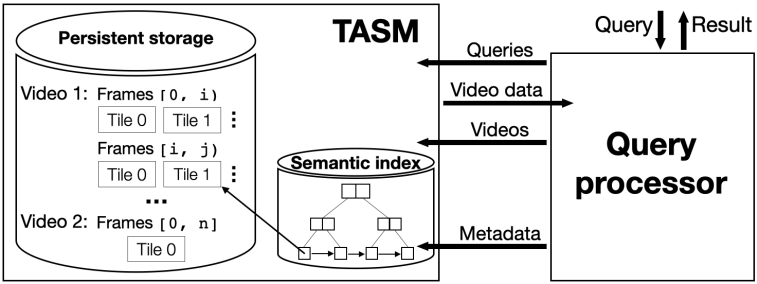

**Figure 7.6:** Architecture diagram of TASM, taken from Daum *et al.* (2021).

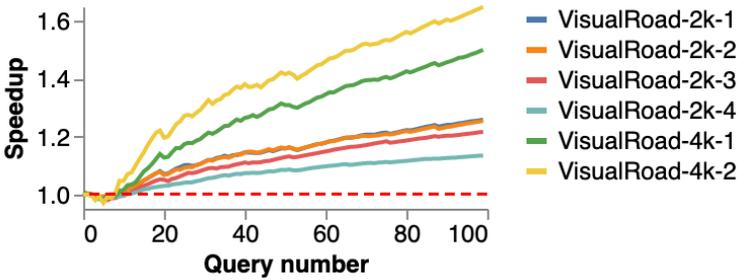

**Figure 7.7:** Speedups of queries when using multiple queries and TASM as the storage layer. Taken from Daum *et al.* (2021).

Now that we have discussed proxies, indexes, and storage, we now turn to efficient query execution on accelerators. These accelerators are typically GPUs.

# 8

# Efficient Query Execution

So far, we have described algorithms for leveraging proxies and constructing indexes. Up to this point, we have not discussed how to efficiently execute these algorithms on hardware. Efficiently using hardware resources is critical for ML-based queries, as it is for standard structured data queries.

Many of the considerations are similar for efficient query execution in standard structured data systems and in unstructured data systems. For example, it is critical to carefully construct systems to leverage all hardware resources available.

However, ML-based queries add another dimension of complexity: there is a direct trade-off between accuracy and computation. This trade-off can affect the choice of proxy model, which can in turn lead to dramatic differences in query execution speeds.

We now turn to describing how these considerations affect high-performance systems for ML-based query execution.

## 8.1 Bottlenecks in Query Execution

One of the first considerations to understand when designing efficient query execution systems is where the bottlenecks to query execution are. For ML-based queries, the bottleneck is nearly always related to the ML model itself.

To understand why, consider the widely used ResNet-50 model for image classification. This model can take up to 4 Gflops to classify a single image (Albanie, 2018). The number of flops dramatically increases with modern large language models (LLMs). For example, a relatively small LLM, Llama2-7b can take up to 1.7 Tflops (Yujie, 2024). Larger models, such as Llama-70b or GPT-3 can take tens to hundreds of Tflops for a single inference pass. In contrast, processing a single structured record could take as few as 100 CPU cycles.

As a result of these costs, nearly all modern ML is performed on the GPU. Thus, studying bottlenecks around the GPU is critical for performance.

Building high-performance ML-based query execution engines requires understanding the entire pipeline for ML inference. This pipeline changes per application, but the general pipeline involves

1. Loading the data off disk,

2. Preprocessing the data on the CPU,

3. Transferring the data to the GPU,

4. Executing the ML model on the GPU,

5. Retrieving and parsing the results.

We show an example of a preprocessing pipeline in Figure 8.1. The preprocessing step varies the most between ML workloads. For text workloads, this can be as simple as turning strings into "tokens" via a mapping algorithm. For vision workloads, this can be as complex as decoding a JPEG image, transforming the image, converting the pixels to `float32`, and normalizing the converted image.

The relative throughput of these different operations varies wildly based on the hardware platform.

One major reason for this variation is the rapid advances in GPU hardware performance. For example, Kang *et al.* (2021c) analyzed the performance of accelerators for performing inference on ResNet-50.

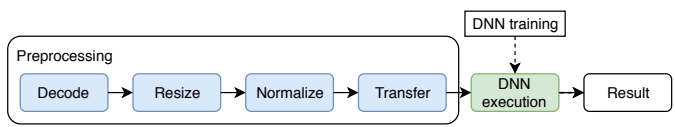

**Figure 8.1:** Example preprocessing pipeline for a vision model. Taken from Kang *et al.* (2021c).

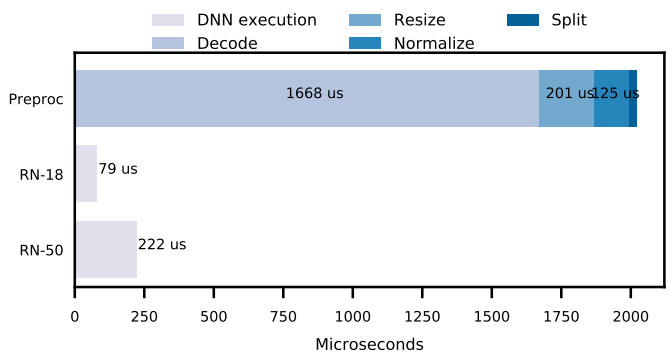

**Figure 8.2:** Breakdown of preprocessing on the CPU compared to inference on the NVIDIA T4 GPU. Inference far outpaces preprocessing, even on older GPUs. Figure taken from Kang *et al.* (2021c).

They found that the throughput increased by 150× in just 6 years. Furthermore, recent accelerators are even higher performance: the H100 can perform up to 55,738 inferences per second (NVIDIA, 2023), which is a nearly 2× improvement over the fastest accelerate considered by Kang *et al.* (2021c). This is a total of a 350× improvement.

CPU performance has largely flatlined. As a result, preprocessing for image workloads can be up to 33× slower than inference. We show an example breakdown of preprocessing vs. inference in Figure 8.2 (taken from Kang *et al.*, 2021c) on the NVIDIA T4 GPU, which is now two generations out of date. These trends have important implications for vision workloads.

In contrast to vision workloads, natural language processing (NLP) tasks appear to be moving towards larger models, with the high performing models having up to 405B parameters. For these workloads, the ML inference is by far the bottleneck.

Now that we have described the characteristics of ML workloads, we now turn to methods of efficient execution of query plans on real hardware.

## 8.2   Efficient Execution on Accelerators

The first class of methods to increase performance are query-agnostic. For example, one method to increase hardware utilization is by batching inputs to the GPU. Because GPUs are massively parallel, they are more efficient when performing inference on batches of data, as opposed to single inputs. Nearly all efficient batch query systems use some form of batching to the GPU (Kang *et al.*, 2021c; Poms *et al.*, 2018; Xu *et al.*, 2019). Other common forms of query-agnostic optimizations include quantization (Liang *et al.*, 2021), efficient use of hardware caches (Kwon *et al.*, 2023), and others (Gou *et al.*, 2021).

The second class of methods to increase performance are query-aware. One major issue is that query-agnostic methods will simply maximize the throughput on the accelerator. But as we have described above, particularly for vision workloads, the cost of preprocessing dominates the cost of small models.

Although small models are not common for the oracle method, they are often used as proxy models. Furthermore, there is a direct trade-off between accuracy and throughput. Thus, choosing the GPU-throughput-optimal model will result in poor accuracy for the proxy model.

To address this, Kang *et al.* (2021c) proposed a system that chooses the most accurate model for preprocessing-bound workloads. For example, consider choosing between a ResNet-18 and ResNet-50 as the proxy model. In this setting, if we only consider the GPU throughput, we would use ResNet-18, which is nearly $3\times$ higher throughput than ResNet-50. However, on modern accelerators, ResNet-18 and ResNet-50 are both preprocessing-bound! Thus, we should always choose ResNet-50 for a system with a modern GPU.

We show the effects of being preprocessing bound in Figure 8.3, which was taken from Kang *et al.* (2021c). As shown, increasing the model size does not improve performance when preprocessing bound.

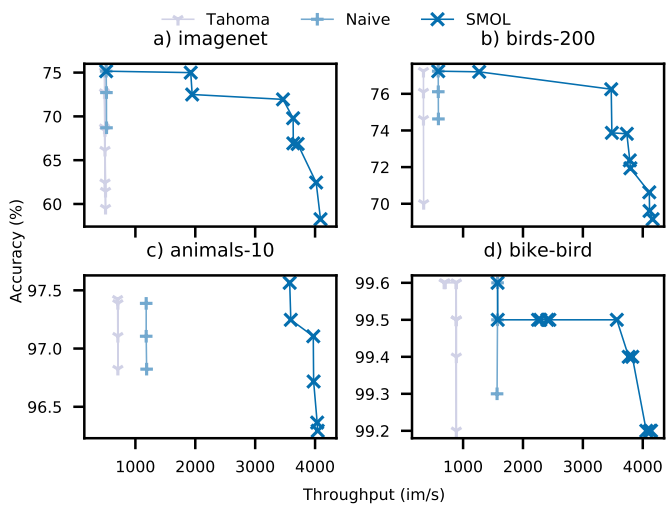

**Figure 8.3:** Accuracy and throughput of Smol compared to baselines that are preprocessing-unaware. As shown, increasing the model size does not improve performance for preprocessing-unaware systems. Taken from Kang *et al.* (2021c).

Another system, Scanner (Poms *et al.*, 2018), focuses on video-specific workloads. Users specify high-level query plans in Scanner, which them compiles them in a massively parallel way. After the query is specified, Scanner will examine the dependency graph to determine where to place the different operations. For example, consider a simplified workload that first decodes a video, then executes an ML model on the decoded frames. Even with this simple workload, one major question is which device to run the video decoding on. If the ML model is expensive, then the decoding and preprocessing should be offloaded to the CPU. If the ML model is cheap, then as much computation as possible should be moved to the GPU.

As we can see from these examples, jointly using the CPU and accelerators is critical to high performance. As accelerators evolve, the systems to support efficient query execution must also evolve. We are excited to see the rise of more advanced systems that efficiently leverage the latest hardware resources.

## 8.3   Local Execution vs. API Execution

One major change over the past few years has been the rise of APIs as a means to interface with ML models. In many circumstances, the weights of highly capable models are not publicly available. For example, the most capable LLM at the time of writing, GPT-4, is only accessible via an API. The cost model of APIs, as with ML model inference, varies for different applications.

The cost is often on a per-image basis for vision tasks. For example, Google Cloud Vision charges \$2.25 per thousand images for object detection at the time of writing. The cost is often on a per character or per token basis for language tasks. OpenAI charges \$5 per million input tokens and \$15 per million output tokens.

Although convenient, it is important to note the cost differentials from using APIs compared to self-hosted ML. For example, Google Cloud Vision charges \$2.25 per thousand images for object detection and localization. In contrast, self-hosting RetinaNet, a state-of-the-art detection model used by MLPerf, can execute at 1,770 samples / second on a single H100 GPU. At a price of \$4.76 / hour / H100, this would cost \$0.00077 per thousand images. It is nearly $3000\times$ cheaper to self-host on a per-hardware cost. Of course, this ignores the price of upkeep, but cost is often critical given the high costs of ML.

Because ML models are typically the overwhelming cost of unstructured data queries, the complexity of query execution systems becomes dramatically reduced. The majority of the effort should go into reducing the number of API calls or token count, as techniques such as batching do not directly apply in this setting. Importantly, offerings from major closed-source LLM providers called "batching" still charge on a per-token basis (although at a reduced rate), so the key determinant to cost is to reduce the total number of tokens used. As such, the techniques we described in prior sections become more important.

# 9

# Video Queries

We now turn to the topic of queries specific to video. Because of the temporal nature of video, many queries require information that span multiple frames. Similar to how standard SQL can struggle with temporal queries, so do row-centric algorithms for answering video queries.

To address this, researchers have built a number of purpose-built systems for answering video queries. We now discuss these systems.

## 9.1 Tracking Queries

The first class of queries we discuss are tracking queries. A tracking query is a query where the user is interested in some information over the trajectory of an object through a video. For example, the user may be interested in counting the number of cars that pass through an intersection. We show an example of a track in Figure 9.1, taken from Bastani *et al.* (2020).

Answering a tracking query inherently requires the ability to associate objects from frame to frame. In computer vision, this is called object tracking (Yilmaz *et al.*, 2006). Object tracking is widely studied in computer vision, as there are many complex scenarios to deal with: multiple objects, occlusion, fast motion, and others.

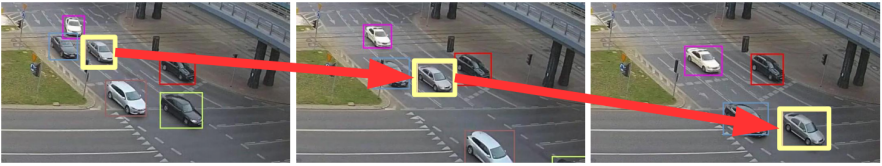

**Figure 9.1:** Example of a track of a car passing through an intersection. An urban planner may be interested in viewing such instances or counting the number of times a track of this form appears in the video. Taken from Bastani *et al.* (2020).

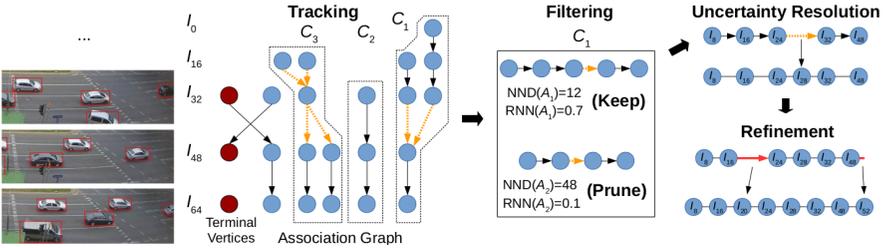

**Figure 9.2:** Architecture diagram for MIRIS. Taken from Bastani *et al.* (2020).

Similar to the naive, exhaustive method for row-oriented queries, tracking queries can be answered by exhaustively executing object detection, object tracking, and custom code on the object tracks. However, this is similarly expensive and can be difficult to express.

Two systems for optimizing tracking queries are MIRIS and OTIF (Bastani *et al.*, 2020; Bastani and Madden, 2022). We show a system diagram of MIRIS in Figure 9.2. Because OTIF is largely a refinement of MIRIS, we focus on OTIF. OTIF exhaustively materializes all of the tracks in a video ahead of time to ensure fast queries at query time.

OTIF refines the exhaustive approach with three components: a frame rate adjuster, a resolution adjuster, and a segmenter based on proxy models. All of these components are designed to decrease the cost of the most expensive part of materializing the tracks: object detection.

The frame rate adjuster selects the frame rate (i.e., drops frames) in the video, so that fewer frames are processed by the object detection method. Fewer frames directly results in lower costs. To understand how dropping frames can retain high quality, consider the example of a traffic camera at an intersection. At night, there may be whole minutes where no vehicles pass through the intersection. In this example, the predictions from the first frame can be carried through.

The resolution adjuster selects the resolution for input for the object detection method (i.e., shrinks frames). The amount of computation of an object detection method is roughly proportional to the number of pixels in the input image for a fixed model, so shrinking frames directly reduces the computational burden. Consider the traffic camera example. If only cars (which are large) are present in a frame, then not much visual fidelity is necessary to detect them. Thus, we can reduce the resolution for many frames in the traffic example.

The segmenter selects parts of the image where the object detection method will be applied. As with the resolution adjuster, fewer pixels passed to the object detection method will result in less computation. Thus, if we can exclude parts of the image, this will reduce computational costs. For static cameras, much of the image is background and can thus be ignored.

Given these three components, the primary question is how to select the parameters (i.e., the frame rate, the resolution, and the segments). This problem is challenging since there are many possible configurations and choices in parameters. Furthermore, the differences in parameters in one setting (e.g., the resolution) will affect differences in others (e.g., the frame rate).

Due to the large number of possible parameters, OTIF solves this problem with hill climbing. We defer details of the optimization procedure to Bastani and Madden (2022).

Other work focuses on optimizing track queries (Chao *et al.*, 2023; Xu *et al.*, 2024) and we refer the reader to these manuscripts for further details.

## 9.2 Actions

Beyond tracks, users are also interested in queries about actions in videos. For example, a user may be interested in finding all occurrences of a car turning left in a video.

Unfortunately, one challenge today is that arbitrary action detection can be low accuracy, even with state-of-the-art models. This is particularly the case for domain-specific tasks, such as analyzing deer feeding patterns. Thus, one of the core challenges for action detection is specifying arbitrary actions. However, even after specifying an action detection model, queries can be expensive to compute using the exhaustive method. We will discuss these two challenges in turn.

### 9.2.1    Specifying Action Queries

One line of work focuses on specifying video queries, of which a subset are action queries. Video queries can be challenge to specify even when given a full schema of object types and locations as standard query languages, like SQL, do not capture the semantics of all user queries. For example, even specifying a query as simple as "car turning left" in SQL can be challenging. We briefly discussed expressing video queries in Section 4.5 and extend our discussion here.

One line of work allows users to specify actions of interest by example (Zhang *et al.*, 2023; Mell *et al.*, 2021). One system, EQUI-VOCAL (Zhang *et al.*, 2023), specifies a scene graph internally, which includes information about objects, their attributes, and their relationship to other objects. It then *synthesizes* queries over the scene graph that match the given examples.

The space of queries is combinatorially large and it is thus challenging to explore the space of all possible queries. This large search space causes two issues: 1) the search algorithm cannot consider all queries and thus must find the relevant queries and 2) many queries can result in the same answers over the limited set of examples.

To resolve these issues, systems that allow for query by example implement a range of strategies, including efficient pruning schemes and active learning. These techniques can also be used in conjunction. At a high level, consider two equally likely queries that match the existing examples. The system can ask the user to label a new example that would distinguish between the two queries. By repeatedly doing this, and by removing queries that are highly unlikely to match, these systems can efficiently synthesize queries from examples. We show an example of EQUI-VOCAL in Figure 9.3.

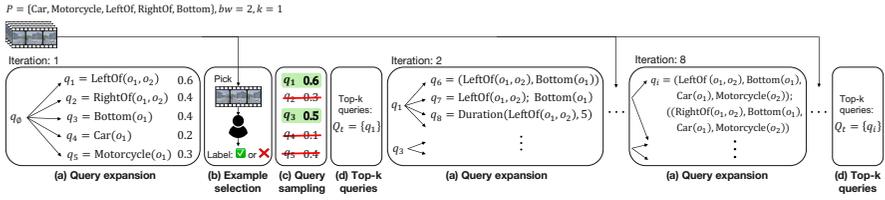

**Figure 9.3:** Example of EQUI-VOCAL, taken from Zhang *et al.* (2023). The system repeatedly selects queries, chooses examples for the user to label, and repeats.

### 9.2.2 Accelerating Action Queries with Known Models

Although the arbitrary action detection problem is challenging, we assume that we are given an action detection model for the problem at hand for the remainder of this section. If trends in AI methods continue, we anticipate that action detection will be "solved" in the next few years, for the common case.

However, this is not yet the case. To answer action queries, existing work assumes that we have some model or method that can identify an action (Chunduri *et al.*, 2022; Chao and Koudas, 2024). For example, the method can be a UDF on top of an object track.

Executing this method exhaustively can be expensive. To address these costs, one system, Zeus, alters the sampling rate, segment length, and resolution for efficient execution (Chunduri *et al.*, 2022). It does so by training a reinforcement learning agent to determine how to tune the three parameters.

Nonetheless, action queries remain an open challenge in the unrestricted case. We are excited to see future research in this area.

### 9.3 DSLs for Video Queries

One active area of research is to define domain-specific languages (DSLs) for video queries. This is because video queries can be extremely challenge to specify using standard SQL due to the ad-hoc and temporal nature of these queries. We have briefly discussed video DSLs in Section 4.5. Two examples are given in Figures 4.2 and 4.4. One DSL, VQPy, is now deployed in Cisco's DeepVision platform.

DSLs allow certain forms of queries to be expressed in a simpler manner compared to standard SQL. For example, finding hit-and-run incidents can be done with fairly few lines of code (Figure 4.4). The developers of the DSL can also implement standard functionality, such as the `SequentialQuery` shown in Figure 4.4. These features are especially useful for temporal queries and action queries.

However, these DSLs still require users to learn a non-standard language. Although VQPy and Rekall are embedding in Python, they require Rekall- and VQPy-specific knowledge. This learning hampers adaption. As we discuss in Section 11, there is a difficult balance between domain-specific systems/languages and general-purpose systems. We are excited to see the future of research in this area.

## 9.4 Other Query Types and Systems

Beyond the specific queries we have discussed, there have been a wide range of video-specific systems built in the past few years (Cao *et al.*, 2022; Canel *et al.*, 2019; Hsieh *et al.*, 2018; Jiang *et al.*, 2018; Krishnan *et al.*, 2018; Moll *et al.*, 2022; Xarchakos and Koudas, 2019; Zhang and Kumar, 2019; Zhong *et al.*, 2023; Xarchakos and Koudas, 2021). These systems range from optimizing fast selection of objects (Canel *et al.*, 2019; Hsieh *et al.*, 2018) to optimizing cross-camera queries (Jiang *et al.*, 2018), among many others with specific forms of video optimizations (Kang *et al.*, 2022b; Chao *et al.*, 2024; Romero *et al.*, 2022; Bang *et al.*, 2023).

# 10

## Text and Semi-structured Queries

We now briefly discuss queries that require processing unstructured or semi-structured text data. These queries are widely varied: they range from processing semi-structured PDFs to open-domain search. Similar to how there is no "settled" way of expressing and processing video queries, there is no single system for processing text data.

To understand why text is so complex, textual data includes from legal opinions (for legal scholarship), police reports (to understand police activities), customer chat support transcripts (to understand business processes), PDFs with semi-structured textual tables (to understand economic or business reports), and many others. Because the type of data and the forms of analyses of interest vary so widely, it is difficult to construct systems are sufficiently expressive to handle all of these cases but restricted enough to not simply be a general-purpose programming language.

Nonetheless, researchers have developed a range of systems and techniques in to begin to address these challenges. As such, we highlight several promising research directions in this area.

## 10.1   Semantic Operators for Text

Consider a simple analysis over Tweets: filter a corpus of Tweets that mention a specific product and compute the sentiment per Tweet. The filtering and sentiment, on a per-Tweet basis, can be accomplished using the map-reduce framework where a text model (such as an LLM) classifies Tweets and computes the sentiment.

The first research direction we discuss is the idea of defining *semantic operators* via text and optimizing them (Patel *et al.*, 2024). A semantic operator is one that transforms one or more relations into other relations that are defined by natural language.

The simplest semantic operator simply maps one column to another. For example, we could map the text of a review to the sentiment of the review. Similarly, we could have two operators to answer the analysis over Tweets above: a filtering semantic operator, and a sentiment semantic operator. The mapping of the sentiment could be executed via an LLM. Beyond maps, we can also construct semantic filters, joins, aggregations, top-k, and group bys. These semantic operators can also be combined with standard relational operators in powerful ways.

We show an example of a pipeline for fact checking defined with semantic operators in Figure 10.1, taken from Patel *et al.* (2024). This pipeline involves a map (to perform relevant search), a semantic join (joining articles and queries), and another map (to classify the purported facts as factual or not). As shown, the semantic operators are defined in natural language and can be chained to create powerful pipelines.

An important question when executing queries with semantic operators is how to optimize such queries. For example, naively executing a top-k operator may result in quadratically many comparisons by performing a pairwise sort algorithm. Instead, we can execute a quick-select algorithm (Hoare, 1961), which is asymptotically faster than a pairwise sort.

Other semantic operators can be optimized as well. For example, we can use a proxy model to accelerate a filter operation by first using a cheaper model.

Semantic operators, when combined with relational operators, can be very powerful. We note the similarity to using UDFs in Eva and

```
1  wiki_df.load_sem_index("article", "index_dir")
2  claim_df.sem_map("write 2 search queries given the {
      claim}", name="query")\
3    .sem_sim_join(wiki_df, left_on="query", right_on="
      articles",  K=10)\
4     # concatenate articles for each claim
5    .groupby(["claim"]).apply(lambda x: "\n".join(x["
      articles"]))\
6    .sem_map("Identify whether there are any factual
      errors in the {claim} based on the {articles}.
      Include your resasoning, any errors found in the
      claim, and the factuality of the claim.")
```

**Figure 10.1:** Fact checking pipeline, using semantic map, sim-join, and map with semantic operators. Taken from Patel *et al.* (2024).

mappings in AIDB–these concepts are related but semantic operators are defined in text. Semantic operators also have similar challenges: efficiency (especially because LLMs are expensive), expressing queries precisely (because natural language is imprecise). We hope that future research addresses these challenges.

## 10.2  Using LLMs to Generate Pipelines

Although semantic operators are powerful, they can be inflexible when handling more complex, semi-structured documents. For example a corpus of documents might contain similar, but slight variations, of structured reports. We would like to have additional flexibility when handling these documents, while maintaining programmability.

To address this, the second direction we consider is LLM-augmented pipelines to answer complex queries. Before we discuss this research direction, we briefly discuss agentic LLM workflows.

As LLMs have improved in capabilities, they now have the ability to write code, perform relatively complex quantitative reasoning, and interact with environments (Masterman *et al.*, 2024). In particular, because they can write their own code, they can use tools like DSLs or Python interpreters. For example, LLM agents can use code to interact with data!

To answer complex queries, we can either manually generate pipelines or use the LLMs themselves to generate these pipelines (typically in an agentic workflow) One such system that enables LLM-augmented pipelines is DocETL (Shankar *et al.*, 2024b). DocETL defines a set of primitives, similar to semantic operators, to transform complex PDF documents. These primitives include maps, filters, and others. These primitives can both be implemented with LLMs, and be executed and rewritten by an agentic LLM workflow.

We show an architecture diagram in Figure 3.2, taken from Shankar *et al.* (2024b). We further show an example of a concrete execution trace in Figure 10.2, also taken from Shankar *et al.* (2024b). As shown, parts of the pipeline can be executed with LLMs and also be rewritten with LLM agents.

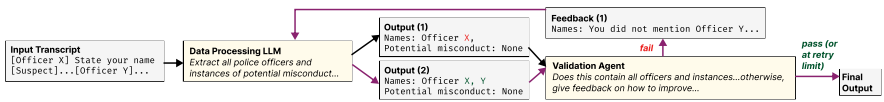

**Figure 10.2:** A concrete execution trace of DocETL, taken from Shankar *et al.* (2024b).

## 10.3   Other Systems

Beyond semantic operators and using LLMs to write pipelines, there is an increasing body of work on efficiently processing text. We highlight several recent systems in this area.

On example of a recent system that uses LLMs is Evaporate (Arora *et al.*, 2023). Evaporate uses either LLMs directly, code synthesis via LLMs, or a mixture to extract attributes from semi-structured documents (e.g., PDFs). The authors find that code synthesis can fail on difficult cases and that optimizing the way LLMs are used results in higher quality and higher performance. Other systems use LLMs or other traditional ML models to create knowledge bases, which can then be queried directly (De Sa *et al.*, 2016).

Another recent system, ELEET, optimizes multi-modal joins (Urban and Binnig, 2024). It does so by extracting key attributes, similar to Evaporate, from text and using the attributes to join. This process is substantially more efficient since using an LLM on the direct multi-modal data requires quadratically many LLM calls, as opposed to extracting attributes.

Finally, we review NeuralDB, which uses natural language to directly query a textual database (Thorne *et al.*, 2021). It operates by creating localized support sets of relevant sentences to answer queries, which enables efficient SPJ operations. However, any such system struggles with aggregation queries, highlighting the need for other work such as semantic operators.

Beyond these two directions, there are many others that aim to allow for ease of expression and optimized text-based queries. The capabilities of AI, particularly LLMs, is rapidly evolving and these research in this area is similarly rapidly changing to enable these capabilities at scale.

# 11

## Open Challenges

Although much work has been done to improve the ease of use and efficiency of unstructured data queries, these systems are not as ubiquitous as structured data systems. However, as we have mentioned, unstructured data volumes far outstrip structured data volumes. Why then are unstructured data systems not as widely deployed?

The answer lies in the two basic questions we have returned to repeatedly in this monograph: expressivity and efficiency. Although seemingly different concerns, they are highly overlapping, as we will discuss.

As we have seen, even expressing query intent to another human can be challenging in some circumstances. Expressing query intent to the computer is even more challenging, given the uncertainty in the query semantics. Even in real-world systems leveraging LLMs, users cannot be guaranteed accuracy semantics, which makes their deployment challenging.

One major trend in research systems is the rise of domain-specific and modality-specific systems. These include video systems (Chao *et al.*, 2020; Zhang *et al.*, 2023; Bastani *et al.*, 2020; Kang *et al.*, 2017), systems for semi-structured documents (Lin *et al.*, 2024), text (Shankar *et al.*, 2024a), and others (Jo and Trummer, 2024). However, domain-specific systems are limited in scope by their nature.

Moving forward, a major question will be whether domain-specific systems will continue to proliferate or whether there will be unified systems for unstructured data.

Beyond expressivity, efficiency is another key concern for unstructured data systems. As we have described, executing an ML algorithm as part of the query execution can result in queries orders of magnitude more expensive than standard structured queries. This cost is a major impediment towards the widespread deployment of unstructured data systems.

Much work has gone towards improving runtimes by trading off accuracy for cost. However, as with all approximate query systems, it can be difficult to users to reason about the trade offs. This is especially true for best-effort systems. As a result, a key challenge is enabling users to understand the trade offs and leverage them effectively to solve their downstream tasks.

Thus, many questions remain around the widespread deployment of unstructured data systems. Will domain-specific systems overcome issues around expressivity and cost while still being broadly applicable? Or will there be unified tooling and interfaces for unstructured data systems? We are excited to see the future of these systems.

# Acknowledgements

We would like to thank the anonymous reviewers for providing comments that we believe improved this manuscript, along with the editors of FnT-DB. We would also like to thank Joe Hellerstein and Surajit Chaudhuri for encouraging us to write this manuscript.